\documentclass[prd,groupedaddress,nofootinbib,showpacs,preprintnumbers,11pt]{revtex4}
\usepackage{epsfig}
\usepackage{amssymb}
\usepackage{amsbsy}
\usepackage{amsmath}

\newcommand{\rf}[1]{(\ref{#1})}
\newcommand{\beq}{\begin{equation}}
\newcommand{\eeq}{\end{equation}}
\newcommand{\be}{\begin{equation}}
\newcommand{\ee}{\end{equation}}
\newcommand{\bea}{\begin{eqnarray}}
\newcommand{\eea}{\end{eqnarray}}
\newcommand{\eq}[1]{Eq.~(\ref{#1})}
\newcommand{\non}{\nonumber \\*}
\newcommand{\ie}{{i.e.}\ }
\newcommand{\bv}{\begin{verbatim}}
\newcommand{\ev}{\end{verbatim}}

\newcommand{\ka}{r}
\renewcommand{\tilde}{\widetilde}

\newcommand{\e}{\,\mbox{e}}
\renewcommand{\d}{{\rm d}}
\renewcommand{\i}{{\rm i}}
\newcommand{\blambda}{\bar\lambda}
\newcommand{\brho}{\bar\rho}

\newcommand{\tr}{\mathrm{tr}}
\newcommand{\LA}{\left\langle}
\newcommand{\RA}{\right\rangle}

\def\lesssim{\mathrel{\mathpalette\fun <}}

\def\fun#1#2{\lower3.6pt\vbox{\baselineskip0pt\lineskip.9pt
\ialign{$\mathsurround=0pt#1\hfil##\hfil$\crcr#2\crcr\sim\crcr}}}

\begin{document}

\preprint{ITEP--TH--34/14}

\title{Scaling behavior of regularized bosonic strings}

\author
{J. Ambj\o rn$\,^{a,b}$ and Y. Makeenko$\,^{a,c}$}

\affiliation{\vspace*{2mm}
${}^a$\/The Niels Bohr Institute, Copenhagen University,
Blegdamsvej 17, DK-2100 Copenhagen, Denmark\\
${}^b$\/IMAPP, Radboud University, Heyendaalseweg 135,
6525 AJ, Nijmegen, The Netherlands\\
${}^c$\/Institute of Theoretical and Experimental Physics,
B. Cheremushkinskaya 25, 117218 Moscow, Russia\\
	\vspace*{1mm}
{email: ambjorn@nbi.dk \ makeenko@nbi.dk}
}


\begin{abstract}
We implement a proper-time UV regularisation of the Nambu-Goto string,
introducing an independent metric tensor and the corresponding 
Lagrange multiplier, and treating them in the mean-field approximation
justified for long strings and/or when the dimension of space-time is large.
We compute the regularised determinant of the 2d Laplacian 
for the closed string winding around a
compact dimension, obtaining in this way the effective action, whose
minimisation determines 
the energy of the string ground state in the mean-field approximation.
We discuss the existence of two scaling limits
when the cutoff is taken to infinity. One scaling limit reproduces  the 
results obtained by the hypercubic regularisation of the Nambu-Goto string as
well as by the use of the dynamical triangulation regularisation of 
the Polyakov string. The other scaling limit reproduces the results obtained by  
canonical quantisation of the Nambu-Goto string.   
\end{abstract}


\maketitle

\section{Introduction}

Recently there has been an increased interest in the spectrum of the large-$N$ QCD
string. It  has been investigated both by numerical simulations
\cite{ABT10,Casele,ABT11,Cas2,AT13,Cas3,AT13a,Cas4,diff,Cas5}
and by analytic studies 
\cite{AF11,AFK12,Dub,Bahns,Vyas,Dub13,AK13,AMS14,Dub14,Dub14a}.
The two major questions to be addressed are: what is the effective action of the QCD
string at large distances and  what is the spectrum of this string?
Addressing the former question implies that we have to modify 
the Nambu--Goto action by adding operators which are less relevant in the 
long-string limit, while the latter question 
requires a consistent quantisation of the string in $D=4$ dimensions.

String theory is generically a nonlinear problem, since the Nambu--Goto action, representing the area of the string world-sheet, is not a quadratic function of 
the fields. However, a gauge fixing makes the action quadratic
with certain constraints imposed on physical states. This is the essence of the canonical
quantisation successfully applied to the relativistic string in the 1970's, which  leads to consistent results in the critical dimension ($D=26$ for the bosonic string) 
and on mass shell.

A subtle feature of the quantized string theory is the emergence of
ultraviolet divergences which have to be regularised  and the theory has to 
be renormalised in order to remove these divergences. In quantum field theory the regularisation is customary done by cutting
off momenta squared above a certain value $\Lambda^2$. In string theory
such a cutoff has to be done for a certain choice of the world-sheet
coordinates (or the choice of  gauge) at $\Lambda^2 \sqrt{g}$, where
$g$ is the determinant of the world-sheet metric tensor,
to comply with  diffeomorphism invariance.
If $g_{ab}$ is the induced metric, 
this may actually result in a complicated nonlinear problem.

The importance of such a dependence of the cutoff on the world-sheet metric is seen
already in the very first computation by Brink and Nielsen \cite{lars-holger}
of the energy due to zero-point fluctuations of 
an open string with fixed ends separated by the distance $L$. The 
classical energy is $E_{\rm cl}=K_0 L$. The energy of zero-point fluctuations is given 
by the sum over the string oscillator modes:
\be
E_0=K_0 \; \frac{D-2}{2} \;\sum_{n=1}^{n_{max}}  \frac n{2E_{\rm cl}} =\frac {D-2}2
\left( \frac{\pi n_{max}^2}{2L} - \frac{\pi}{12L} \right).
\ee
The universal  (\ie regularisation-independent) second term on the right-hand side comes as  the difference between the actual sum of  discrete modes and 
an integral approximation to the sum.
Diffeomorphism invariance requires that
the maximal number of modes, $n_{max}$, is $L$-dependent: 
$ n_{max} =  L \Lambda/\pi$.
We thus obtain
\be
E_0=(D-2) \left(\frac{\Lambda^2 L}{4\pi}  - \frac{\pi}{24L} \right).
\label{zpe}
\ee
Therefore the divergence contributes only
to the string tension, but not to the lowest mass, which is determined by
the (universal) second term on the right hand side, whose negative sign is
associated with a tachyon.
If we naively used $ n_{max} = {\rm const}$, it would result
in a divergent (and positive) mass squared of the lowest state of the string.

The dependence of the cutoff on the world-sheet metric is crucial for
the path-integral  formulation of string theory~\cite{Pol81},
where the world-sheet metric $g_{ab}$ 
and the target-space position $X^\mu$,
$\mu =1,\ldots,D$,
of the string world-sheet are independent.
Owing to diffeomorphism invariance, the world-sheet 
metric can be 
diagonalized, $g_{ab}=\e^\varphi \delta_{ab}$, by choosing the conformal gauge.
While the classical action does not depend on $\varphi$, it emerges in the 
effective action after the path-integration over  $X^\mu$ (and ghosts)
because of the divergences of these path integrals and the corresponding 
dependence of the cutoff on $\varphi$. However, the remaining  
path integral over $\varphi$ decouples  on mass shell in 
the critical dimension, and then the results  obtained in the 1970's 
using the operator formalism are reproduced.
For $D\neq26$ and/or off shell, the path integral over $\varphi$ 
has to be taken into account and plays an important role for the consistency. 

In the Polyakov formulation of string theory
the path integral over the target-space string coordinates 
(and ghosts) 
is Gaussian and results in the determinant of the 
2d Laplace-Beltrami operator with proper boundary conditions.
For an open string with fixed ends these are the
Dirichlet boundary conditions, for which the computation 
of the determinant was performed in \cite{DOP82,Alv83}. 
For slowly varying fields $\varphi$, the effective action 
is determined by the conformal anomaly and given by the 
so-called Liouville action. 
Remarkably, the path integral over $\varphi$
can be consistently treated~\cite{AMS14} order by order in 
the inverse string length and/or in the limit of large  $D$, 
where the WKB expansion about 
the saddle points applies. 
The result for the
ground-state energy as a function of $L$
coincides with the well-known Alvarez-Arvis spectrum~\cite{Alv81,Arv83}.
It reveals a tachyonic singularity at distances $L\leq L_0$ 
with $- K_0^2 L_0^2$ being the tachyon mass squared.
For larger distances this quantity is well-behaved.

The Alvarez-Arvis spectrum follows \cite{Arv83,Ole85} from the canonical quantisation of an open string 
with the Dirichlet boundary condition. Similarly, no effect of nonlinearities are  seen
in the computation \cite{Alv81} 
of the Nambu--Goto path integral at large $D$ when one uses  the
zeta-function regularisation.  However, one may wonder 
whether this regularisation always can be used since a power-like divergence
is missing by construction. It is thus not obvious to which extent the 
dependence of  the cutoff on the metric, which was the origin of 
the non-linearity, is correctly captured using the zeta-function regularisation.
For this reason we would like to repeat the computation using a regularisation 
where the UV cutoff is given by a dimensional  parameter, like the Pauli-Villars or proper-time regularisation. Without such a dimensional parameter it is hard to follow 
how the regularisation affects the renormalisation of the string tension 
or the masses in the theory.

There exist two lattice-like string theories%
\footnote{For an introduction see the book \cite{ADJ97}.}
where the UV cutoff is explicitly a dimensional 
parameter, the lattice length $a\sim1/\Lambda$. In the first lattice approach the 
starting point is the Nambu-Goto action, 
and the path integral over string configurations  is regularised by
considering surfaces embedded on a hypercubic lattice, living on the plaquettes of 
the lattice. We denote this theory the hypercubic lattice string theory (HLS)
\cite{lattice}. 
The second lattice theory is a regularisation of the Polyakov string theory,
 often denoted dynamical triangulation (DT)~\cite{DT},
since summation over the intrinsic geometries of the world-sheet in the path integral is performed by summing over a suitable class
of equilateral triangulations, each with link length $a$. The target space variables 
$X^\mu$ then live on the vertices of the triangulations and in this way a target space 
triangulation is defined. Both theories are naturally defined in any Euclidean 
$R^D$ target space, where $D \geq 2$, 
and both theories lead to the following picture:
one can define a two-point function which falls off exponentially at large target
space distances, thus defining a renormalised lowest, positive mass of the theories.
However, once this two-point function  is defined, the string tension
of the theory does not scale~\cite{lattice,AD87}, 
i.e.\ it goes to infinity when the cutoff $a\to 0$.
Thus, these regularised theories seemingly had little to do with 
ordinary bosonic string theory. 

However, so-called non-critical string theory,
which can be viewed as an extension of string theory to $D < 2$, showed that
the DT-regularisation indeed captured precisely what one would view as
string theory in this region (it is not known how to extend  HLS to $D<2$).
Since bosonic string theory ceases to be tachyonic for $D<2$, the problem for
the lattice versions of string theory seems to be the fact that by construction
they have no tachyons. From their very construction, the logarithm of the 
two-point function in these theories is sub-additive, leading to a mass 
larger than or equal zero (see for instance \cite{ADJ97} for a discussion). 

Clearly the continuum bosonic string theory manages to perform subtractions which 
result in a negative $m^2$ (and a finite, positive renormalised string tension), but 
it has never been fully understood how to reconcile the continuum calculations
with the lattice calculations where it seems plain impossible to obtain a negative
$m^2$. In order to avoid this conundrum we will try to put ourselves in 
a string theory situation where there is no tachyon in the continuum 
formulation, where $D>2$ and where there is thus a chance that the lattice 
and the continuum formulations might agree, precisely as they agree 
in the case $D < 2$, as mentioned above.  

Remarkably, such a comparison
is possible in the large $D$ limit if we consider a closed string which propagates
a distance $L$ and which is wrapped around  one target space dimension 
compactified to a circle with circumference  $\beta \ll L$.  For $\beta$ not too
small there is no tachyon according the old calculation by Alvarez~\cite{Alv81}. 
We will repeat the calculation, using the Nambu-Goto action and a Lagrange multiplier 
for the induced world-sheet metric. In the large $D$ limit a saddle point 
calculation is reliable and we will perform the calculation using a proper time 
cutoff $a$ which serves as the equivalent to the lattice cutoff mentioned above.
We will find a remarkable situation: one can follow the philosophy of the lattice
approach and first renormalise  the two-point function. It leads naturally to 
a certain renormalisation of the string tension. However, this renormalisation is 
then incompatible with a finite effective action for an  ``extended'' string where
$L \gg \beta$, and where $\beta \gg a$. From the perspective of the effective action
of the extended string it implies that the effective string tension goes to infinity
when the cutoff $a \to 0$, i.e.\ precisely the situation encountered for the lattice
regularisations. However, due to a scale invariance of the effective action, it 
is in the continuum possible to avoid this situation by a rescaling of the 
target space coordinates, but the price one pays is the introduction of 
the tachyon.

The rest of this article is organized as follows: In Sect.\ \ref{RW} we consider
the  particle, rather than the string, using the length
of the world-line  as the action, the particle equivalent of the Nambu-Goto action for
 the  string. The technique and many of the results are the same as for the string,
just simpler.
In Sect.~\ref{s:NG} we consider a path-integral formulation of the Nambu-Goto string,
introducing an independent metric tensor and the corresponding Lagrange multiplier.
In Sect.~\ref{s:s.p.} we find a saddle-point solution to
the path-integral formulation of the Nambu-Goto string
which is justified by the mean-field approximation and
becomes exact at large $D$.
The dependence of the ground-state energy on the string length $L$ is
computed in Sect.~\ref{s:ge}. We also evaluate there the value of the area of 
typical surfaces which dominate the path integral.
Our main results concerning two possible scaling behaviors are presented
in Sect.~\ref{s:ge} and \ref{standardS}.
In Sect.~\ref{s:Pol} we show how the same results can be obtained using the
Polyakov string formulation. The spectrum of excited states is briefly discussed
in Sect.~\ref{s:exa}. Our main results are summarized in Sect.~\ref{Conclusion}.
In Appendix~\ref{AppB} we remind the reader of 
some results for path integral of  the relativistic particle.
In Appendix~\ref{appA} we compute the induced metric for 
the string and find its
unexpected coordinate dependence near the boundary.

\section{Sum over paths for the relativistic particle\label{RW}}

Before performing the calculation for the bosonic string it is 
instructive to consider a similar calculation for the relativistic 
particle, using the length of the world-line as the action, the 
particle equivalent of the Nambu-Goto action we will use for the string.

We rewrite the action (the bare mass times the length of the path) as
\be\label{B1}
S=m_0 \int \d \omega\, \sqrt {\dot x ^2}=m_0 \int \d \omega\, \sqrt{h} +
\frac{m_0}2 \int \d \omega\, \lambda \left(\dot x^2 -h \right),
\ee
where $\dot{x} = d x(\omega)/d\omega$ and where we have 
introduced an independent world-line metric $h$ which is 
a tensor [$h(\omega) = h_{11}(\omega)]$, and a Lagrange multiplier 
$\lambda=\alpha/\sqrt{h}$ with $\alpha$ being a scalar.

The classical equations of motion are
\begin{subequations}
\bea
\frac1{\sqrt{h}}\frac{ \d } {\d \omega} \lambda \dot x^\mu &=&0, \\*
h&=&\dot x^2,\\* \lambda &=&\frac 1{\sqrt{h}}.
\eea
\end{subequations}
A generic solution is
\be
x_{\rm cl}^1(\omega)=\int_0^\omega \d\omega' \sqrt{h_{\rm cl}(\omega')},\quad
x_{\rm cl}^2=\ldots=x_{\rm cl}^D=0.
\label{genera}
\ee
We can choose the (static) gauge, where 
\be
x_{\rm cl}^1= \frac{\omega}{\omega_L} L,\quad \sqrt{h_{\rm cl}}= \frac{1}{\lambda} = 
\frac{L}{\omega_L},  \quad \omega\in
\left[0,\omega_L \right]
\label{gaugep}
\ee
and $L$ is the distance in the target space between the endpoints of the path.
Of course any change of parametrization $\omega \to \omega'(\omega)$ will
also provide us with a classical  solution $x'(\omega')=x(\omega)$ and 
$h'_{11}(\omega') (d\omega')^2 =h_{11}(\omega)(d\omega)^2$.

Splitting $x^\mu=x_{\rm cl}^\mu +x_{\rm q}^\mu$ where  $x_{\rm cl}^\mu $ is
given by \rf{genera} and then 
integrating over $x_{\rm q}^\mu$, we find the effective action
\be
S_{\rm eff}=m_0 \int \d \omega\, \sqrt{h} +
\frac{m_0}2 \int \d \omega\, \lambda \left(\dot x_{\rm cl}^2 -h \right)
-d \Lambda  \int \d \omega \frac{\sqrt[4]{h}}{\sqrt{\lambda} }
+ \frac{d }2 \log \left(\Lambda{ \int \d \omega \frac{\sqrt[4]{h}}{\sqrt{\lambda}}}
\right).
\label{Seffp}
\ee
More specifically the integration over $x_{\rm q}^\mu$ results in the determinant of the 
diffeomorphism invariant differential operator 
\beq\label{janx1}
{\cal O}  = -\frac 1{\sqrt{h}} \frac{\d}{\d\omega} \lambda  \frac{\d}{\d\omega}. 
\eeq
We regularise this determinant by using a proper time cutoff
\be
\tr\log {\cal O} 
= - \int_{a^2}^\infty \frac{\d\tau}\tau \tr  \e^{-\tau {\cal O}}, \quad
a^2 \equiv \frac1{4\pi \Lambda^2 } ,
\label{5}
\ee
By an  explicit calculation for constant  $h$ and $\lambda$ we obtain
\be\label{janx2}
\tr \log\left( -\frac 1{\sqrt{h}} \frac{\d}{\d\omega} \lambda  \frac{\d}{\d\omega} \right)
= -\int_{a^2}^\infty \frac{\d\tau}\tau
\sum_{n=1}^\infty \exp{\left[ -\frac\tau{\sqrt{h}} \lambda \left( \frac{\pi n}{\omega_L}\right)^2 \right]}=-\frac{\omega_L}{\sqrt{\pi}a } \frac{\sqrt[4]{h}}{\sqrt{\lambda} }+\log \frac{\omega_L\sqrt[4]{h}}{a \sqrt{\lambda}},
\ee
which finally leads  to \rf{Seffp}.

In addition to the path integral over $x^\mu$ which resulted in the effective 
action \rf{Seffp}, we have path integrals over the fields $h$ and $\lambda$. 
As is well known,%
\footnote{See, e.g.\ the book~\cite{Pol87}, Sect.~9.1.}
 the path integral over $\lambda$ is saturated by a constant
value of  $\alpha$ owing to localisation, 
after which the dependence on $h$ enters only via the
length $\tau=\int \d\omega \sqrt{h}$ of the path.
The path integral over $h$ (factorised over reparametrisations $f(\omega)$, 
$f^\prime(\omega)\geq 0$, of the path) 
 can then be  substituted by an ordinary integral over $\tau$ 
\be
\int \frac{{\cal D}h}{{\cal D}f} \cdots = \int \frac{\d \tau}{\sqrt{\tau}}
\det{}^{1/2} \left(- \frac 1{\sqrt{h}} \frac{\d}{\d\omega}  \frac 1{\sqrt{h}} \frac{\d}{\d\omega}
\right)\cdots.
\label{jjjjj}
\ee
This will change $D\to D-1$ in the linear divergence of the effective action in
accordance with the fact that there are only $D-1$ independent degrees of freedom
for the relativistic path.

In the rest of this section we shall ignore such a shift of $D$ assuming that $D$
is large. For the string the shift will be from $D$ to $D-2$. We will use the 
notation $d$ for the shifted value of $D$ with the understanding that 
it makes not difference in the large $D$ limit. 
For $m_0\sim d$ all the terms in the action \rf{Seffp} would be of
order $d$, so the Jacobian displayed in \eq{jjjjj} will be not essential.
We can then compute the integral over $\tau$ by the saddle-point method.
Equivalently, we can simply compute the path integrals over $h$ and $\lambda$
at large $d$ by the saddle-point method, minimising the effective action \rf{Seffp},
without introducing the variable $\tau$. This is exactly how we shall proceed in
the next sections when we deal with the relativistic string.

Minimising \rf{Seffp} with respect to $h$, we obtain the equation for $\alpha$
\be
1-\alpha-\frac{d \Lambda}{2 m_0\sqrt{\alpha}}
+\frac{d}{4m_0\sqrt{\alpha}\int\d\omega \sqrt{h/\alpha}}=0.
\label{B4}
\ee
The solution is an $\omega$-independent constant. Since we can always 
choose $h$ to be constant in one dimension by change of parametrization, 
this shows that it is not inconsistent to choose both $h$ and $\lambda$ constant,
as was done in the calculation \rf{janx2}.

Minimising \rf{Seffp} with respect to $\lambda$, we obtain the equation for $h$
\be
\dot x_{\rm cl}^2 -h +\frac{d \Lambda}{m_0\alpha^{3/2}} h 
-\frac{d h}{2 m_0\alpha^{3/2}\int\d\omega \sqrt{h/\alpha} }  =0 ,
\label{hhhhh}
\ee
relating $h$ and $x_{\rm cl}$. 
 Using \eq{B4}, we write
\be\label{janx4}
h=\frac\alpha {(3\alpha-2)}\dot x_{\rm cl}^2  = 
\frac{\alpha}{(3\alpha-2)} \frac{L^2}{\omega_L^2}.
\ee
From \rf{B4} it follows that the ``bare'' mass 
$m_0$ has to diverge as $\Lambda$ for $\Lambda \to \infty$.

At the minimum, we have the following leading large $L$ behavior
\be
S_{\rm eff}=m_0\left( 3\alpha-2\right)\int \d \omega\, \sqrt{h} = 
m_0\sqrt{\alpha\left( 3\alpha-2\right)} L.
\label{Seffp*}
\ee
Since our effective action is just the logarithm of the free particle propagator,
we know that the leading  $L$ behavior is 
\beq\label{janx3}
S_{\rm eff} = m_{\rm ph} L + {\cal O}(\log L)
\eeq
where $m_{\rm ph}$ is the physical mass of the particle. This is obtained by 
choosing  
\be
\alpha = \frac23 + \frac{m_{\rm ph}^2}{2m_0^2}.
\label{B13}
\ee
Equation~\rf{B4} then says
\be
m_0=\sqrt{\frac{27}{8}} d\Lambda +\sqrt{\frac32} \frac{m_{\rm ph}^2}{d\Lambda},
\label{B140}
\ee
the scaling relation well known from treating the relativistic particle as 
a limit of a random walk process with average step length $a \sim 1/\Lambda$
(\cite{ADJ97}). The value $\alpha = 2/3$ is far away from the classical 
value $\alpha =1$. However, Eqs.\ \rf{B4} and \rf{hhhhh} have semiclassical power expansions in $d\Lambda /m_0$ ($ 1/m_0 \propto \hbar$), starting out with 
$\alpha =1$ and decreasing towards $\alpha = 2/3$ with decreasing $m_0$.
The radius of convergence of this expansion corresponds precisely
to $\alpha = 2/3$, as is shown in Appendix~\ref{AppB},
and the value $m_0=d\Lambda \sqrt{27/8}$ associated with
$\alpha =2/3$ is thus the natural quantum point 
of the free particle. As we will see the situation will be similar for the string.

Classically the length of the particle path is $\int \d \omega \sqrt{h_{\rm cl}} =L$.
However, the average path in the path integral is much longer, as is clear 
from \rf{janx4}, which shows that the average length of such a path is 
\beq\label{janx5}
\ell = \LA \int \d \omega \sqrt {\dot x^2} \RA=
 \int \d \omega  \sqrt{h} =
\sqrt{\frac{\alpha}{3\alpha-2} }L.
\eeq
The reason for the divergence of $\ell$ when the cutoff $a \to 0$ is 
of course the quantum fluctuations of $x_q$. One can explicitly calculate 
(see Appendix~\ref{AppB}) that in the limit where $a \to 0$ we have 
\be
\LA \dot x_q ^2 \RA =
\frac{d \Lambda \sqrt[4]{h}}{m_0\lambda^{3/2}} = h.
\label{B14}
\ee

Equation~\rf{janx5} shows that the Hausdorff dimension of  such a path is 
two in the scaling limit and that the proper time cutoff $a$, even if introduced
as a diffeomorphism invariant cutoff in parameter space
$\omega \in [0,\omega_L]$, has a consistent interpretation as a 
length $a$ in target space. Let us assume that $a$ can be interpreted 
as a typical smallest length scale probed in target space. 
Then we can view the path of length $\ell$ as made of $n_\ell = \ell/a$ 
pieces or ``building block''. 
Similarly the classical path $x_{\rm cl}$ consists of $n_L = L/a$ building blocks
and \rf{janx5} reads in the scaling limit:
\beq\label{janx7}
n_\ell = \sqrt{\frac{3}{8\pi}} \;  \frac{d}{m_{\rm ph} L}  \, n_L^2,
\eeq
which tells us that the path of length $\ell$ with endpoints separated by a distance 
$L$ in target space has Hausdorff dimension $d_H=2$. 

We remind the reader 
that there is nothing wrong with the result that the average length of 
a path appearing in the path integral diverges when the cutoff is removed.
As is well known, even in ordinary quantum mechanics, 
such a path is not an observable.
In the Heisenberg picture the operators $\hat{x}^\mu(t)$ 
do not commute at different times and  attempts to measure $\hat{x}^\mu(t )$ at successive small time
intervals $\Delta t$ will precisely result in an average fractal path with 
$d_H=2$ in the limit $\Delta t \to 0$.

The fact that the Hausdorff dimension $d_H=2$ is linked to the interpretation of 
the proper time cutoff $a$  as a length scale also
in target space. From the explicit expressions \rf{janx1} and \rf{5} it is 
clear that in the classical limit where $\lambda /\sqrt{h} = (\omega_L /L)^2$
oscillating modes with mode number $n > L/\pi a$ will be suppressed, telling
us that we can probe distances down to $a$ in target space by the fluctuating field
$x^\mu$. However, from \rf{janx5}  the corresponding mode cutoff in the 
quantum case is $n > \ell/\pi a \propto L/ m_{\rm ph} a^2$.  The fact that the 
path has length $\ell \gg L$ implies that we have to use a much larger 
frequency $\omega$ when expanding $x^\mu$ in modes in order to 
obtain the same resolution in target space.

It is possible to perform a different scaling. Suppose we insist that $\ell$ is 
finite. This is what one would do if we considered one-dimensional 
gravity and $x^\mu(\omega)$ were fields living in this one-dimensional 
world%
\footnote{Of course the gravity formulation would be even clearer if 
we had used to Brink-Howe-DiVecchia formulation, where we have
an independent metric $h_{11}(\omega)$ and a free Gaussian field
$x^\mu(\omega)$ coupled covariantly to $h_{11}(\omega)$, i.e. the 
particle equivalent of the Polyakov string formulation. However, the 
results will be the same as the ones we have already derived, so 
we will refrain from giving any details in the case of the particle. For the 
string we will consider the Polyakov formulation in addition to the Nambu-Goto
formulation.}. In such a world we expect the leading term in the
effective action \rf{Seffp*} to be
proportional to the one-dimensional volume $\ell = \int d\omega \sqrt{h}$, 
i.e. one would write
\beq\label{janz1}
S_{\rm eff} = \tilde{m}_{\rm ph} \ell, \quad   \tilde{m}_{\rm ph} = m_0 (3\alpha -2),
\eeq
or
\beq\label{janz2}
\alpha = \frac{2}{3} + \frac{\tilde{m}_{\rm ph}}{3m_0}
\eeq
and
\be
m_0= \sqrt{\frac{27}{8}} d\Lambda+ \sqrt{\frac{2}{3}}\tilde {m}_{\rm ph}
\label{B141}
\ee
instead of the scaling \rf{B13}, \rf{B140}. From the perspective of such a one-dimensional
world a finite $\ell$ implies that our  former target space $L$ is as small as
the cutoff $a \sim 1/\Lambda$. However, from the viewpoint of our one-dimensional
world $x^\mu$ is just a field and in the split $x = x_{\rm cl} + x_q$, where integration 
over quantum $x_q$ produces the different scaling of $L$ and $\ell$, we
are free to perform a renormalisation of the background field 
\beq\label{janz3}
x_{\rm cl} = Z^{1/2} x_R, \quad Z = (3\alpha -2)/\alpha.
\eeq
The field renormalisation $Z$ has a standard perturbative expansion
\beq\label{janz4}
Z = 1 -  m_0^{-1} \; d\Lambda +{\cal O}(m_0^{-2}),
\eeq
in terms of the coupling constant $m_0^{-1}$, which in perturbation theory
is always assumed to be small even compared to the cutoff. By such a renormalisation
we obtain a new $L_R$ in target space, $L = Z^{1/2}L_R$, which scales the 
same way as $\ell$ and the effective action is simply changed from 
$m_{\rm ph} L$ to $ \tilde{m}_{\rm ph} L_R$ (and a complete calculation
of effective action which also includes power corrections will lead to 
identical expressions, except for an allover, cutoff dependent normalisation factor). 

We will see that similar relations are valid for the Nambu-Goto string, but they 
will have more radical consequences in the string universe.

\section{The Nambu-Goto string\label{s:NG}}

We now use the Nambu-Goto action and perform a calculation
similar to the one for the particle. 
This was first done by  Alvarez~\cite{Alv81} at large $D$ and 
extended by Pisarski and Alvarez~\cite{PA82} to the topology of a cylinder.
As described in the introduction
the set up is the following: we have a closed string propagating
a distance $L$ and wrapped around a compactified dimension of 
circumference $\beta$.  
The action is diffeomorphism invariant and the results should not depend on the 
chosen parametrization.

Introducing an auxiliary field $\lambda^{ab}$ and independent metric field $\rho_{ab}$,
we rewrite the Nambu-Goto action in the standard way as
\be
K_0 \int \d^2\omega\,\sqrt{\det \partial_a X \cdot \partial_bX}=
K_0 \int \d^2\omega\,\sqrt{ \det\rho_{ab}} 
+\frac{K_0}2 \int \d^2\omega\, \lambda^{ab} \left( \partial_a X \cdot \partial_bX -\rho_{ab}
\right).
\label{aux}
\ee
Here $\lambda^{ab}$ transforms under coordinate transformations as a tensor
times the volume element and $\rho_{ab}$ is a tensor. 
The path integration is performed independently  over real values of $X^\mu$ and $\rho_{ab}$ and over imaginary values of $\lambda^{ab}$.

The Euler-Lagrange equations, minimising the right-hand side of \eq{aux} 
with respect to $X^\mu$, $\lambda^{ab}$ and $\rho_{ab}$ are
\begin{subequations}
\bea
&&\frac1 {\sqrt{\det{\rho}}}\partial_a \lambda^{ab}\partial_b X^\mu=0 , \label{Xc} \\*
&&\rho_{ab}= \partial_a X\cdot \partial _b X, \label{rhoX} \\*
&& \lambda^{ab}= \rho^{ab} \sqrt{\det{\rho}},\quad \det{\rho}\equiv\det{\rho_{ab}} .
\label{rl}
\eea
\label{E-L}
\end{subequations}
Choosing the world-sheet parametrization with $\omega_1$ and $\omega_2$ inside 
a $\omega_\beta\times \omega_L$ rectangle in the parameter space, we find from \eq{E-L}
\begin{subequations}
\bea
&&X^1_{\rm cl}=\frac L {\omega_L} \omega_1,\quad
 X^2_{\rm cl}=\frac\beta {\omega_\beta} \omega_2,\quad
 X^\perp _{\rm cl} =0,\label{Xcla} \\*
 && 
\left[ \rho_{ab} \right]_{\rm cl}={\rm diag}\left(\frac{L^2}{\omega_L^2},\frac{\beta^2}{\omega_\beta^2}\right),
\label{rhocla} \\*
&& \lambda^{ab}_{\rm cl}={\rm diag}\left(
\frac{\beta\omega_L}{L\omega_\beta},\frac{L\omega_\beta}{\beta\omega_L}\right).
\label{lacla}
\eea
\label{cla}
\end{subequations}

To analyze quantum fluctuations in the path-integral approach, it is convenient to split 
$X^\mu=X^\mu_{\rm cl} +X^\mu_{\rm q}$, where  $X^\mu_{\rm cl} $ is
given by \eq{Xcla}, and perform the Gaussian path
integral over $X^\mu_{\rm q}$. 
We may fix the gauge at this stage, e.g.\ by choosing $X_{\rm q}^1=X_{\rm q}^2=0$,
\ie choosing the so-called  static gauge,
where fluctuations are transversal to
the classical string world-sheet.%
\footnote{Fixing a static gauge produces a ghost determinant, which is a determinant of
an operator of multiplication by a function. At large $D$ this determinant can be
ignored, but may become essential to next orders of the $1/D$-expansion.}
The number of fluctuating $X$'s then equals
the number of dimensions transversal to the string world-sheet: $d=D-2$. 
We then obtain the effective action,
governing the fields $\lambda^{ab}$ and $\rho_{ab}$
\be
S_{\rm eff}= K_0 \int \d^2\omega\,\sqrt{ \det\rho} 
+\frac{K_0}2 \int \d^2\omega\, \lambda^{ab} \left( \partial_a X_{\rm cl } \cdot \partial_bX_{\rm cl } -\rho_{ab}
\right)+ \frac d2 \tr \log \left( -\frac1 {\sqrt{\det{\rho}} }  \partial _a \lambda^{ab} \partial_b  \right),
\label{aux1}
\ee
where $d=D-2$ is the number of fluctuating $X$'s.

We use the proper-time regularisation of the trace as in \rf{5}, now with
\be
{\cal O}=-\frac{1}{\sqrt{\rho}} \partial _a \lambda^{ab} \partial_b,
\quad \sqrt{\rho}\equiv \sqrt{\det{\rho}} ,
\label{calO}
\ee
which reproduces the usual 2d Laplacian for  $\lambda^{ab}$ given by \eq{rl}.

Using the invariance of the measure in the path integral over
$X^\mu$, $\lambda^{ab}$ and $\rho_{ab}$, we derive the following 
exact set of the quantum Schwinger-Dyson equations
\begin{subequations}
\bea
&&\LA F\left[ \lambda,\rho \right]
\frac1{\sqrt{\rho}}\partial_a \lambda^{ab}\partial_b X^\mu_{\rm cl}\RA =0
, \label{Xcq} \\*
&&\LA \rho_{ab}F\left[ \lambda,\rho \right]\RA= 
\LA \partial_a X\cdot \partial _b X F\left[ \lambda,\rho \right]\RA
+\LA\frac1{K_0} \frac{\delta F\left[ \lambda,\rho \right]}
{\delta \lambda} \RA, \label{rhoXq} \\*
&& \LA \frac{\lambda^{ab}}{\sqrt{\rho}} F\left[ \lambda,\rho \right]\RA= 
\LA \rho^{ab}\left(1-\frac {d}{2K_0\sqrt{\rho}}
\langle \omega | \e^{-a^2 {\cal O}}|\omega\rangle\right) F\left[ \lambda,\rho \right]\RA+\LA \frac1{K_0 \sqrt{\rho}}
\frac{\delta F\left[ \lambda,\rho \right]}
{\delta \rho_{ab}}\RA.
\label{rlq}
\eea
\label{S-D}
\end{subequations}
Here $F\left[ \lambda,\rho \right]$ is an arbitrary functional of $\lambda^{ab}$
and $\rho_{ab}$.

In the mean-field approximation, which becomes exact at large $D$, 
we can disregard fluctuations
of  $\lambda^{ab}$ and $\rho_{ab}$ around the  saddle-point values,
\ie simply substitute them by mean values. 
This is analogous to what happens in the
$N$-component sigma-model at large $N$, where we can disregard quantum
fluctuations of the Lagrange multiplier.

Disregarding the quantum fluctuations means 
 in the path-integral language that the path integrals over
$\lambda^{ab}$ and $\rho_{ab}$ are given by saddle points.
These saddle points can be alternatively found from the whole set of 
the Schwinger-Dyson equations \rf{S-D}, assuming factorisation.%
\footnote{See e.g.\ p.\ 247 of the book \cite{Mak02}.}
The Schwinger-Dyson equations are then reduced to three 
equations for the saddle-point values
$\blambda^{ab}\equiv \LA \lambda^{ab}\RA$ and $\brho_{ab}\equiv\LA \rho_{ab} \RA$:
\begin{subequations}
\bea
&&
\frac1{\sqrt{\brho}}\partial_a \blambda^{ab}\partial_b X^\mu_{\rm cl}=0
, \label{Xcqsp} \\*
&& \brho_{ab}= 
\LA \partial_a X\cdot \partial _b X \RA
, \label{rhoXqsp} \\*
&& \frac{\blambda^{ab}}{\sqrt{\brho}}= 
\brho^{ab}\left(1-\frac {d}{2K_0\sqrt{\brho}}
\langle \omega | \e^{-a^2 {\cal O}}|\omega\rangle\right) .
\label{rlqsp}
\eea
\label{spp}
\end{subequations}
Equations \rf{Xcqsp} and \rf{rhoXqsp} look  similar to 
the classical Eqs.~\rf{Xc} and \rf{rhoX}, while \eq{rlqsp}
contains an additional term compared to the classical \eq{rl},  due to the fact that
operator ${\cal O}$ in \eq{calO} depends explicitly  on $\rho$.

Using the known Seeley expansion for the cylinder, we write \eq{rlqsp} in the bulk
(\ie away from the boundary)  as
\be
\blambda^{ab}= 
 \brho^{ab}\sqrt{\brho}\left(1-\frac {d \Lambda^2 }{2K_0\sqrt{\det \blambda^{ab}}}
\right).
\label{rlqS}
\ee
This equation possesses the solution
\be
\blambda^{ab} = C\brho^{ab}\sqrt{\brho},\quad C=\frac12 +\sqrt{\frac14-
\frac{d\Lambda^2}{2K_0}},
\label{laws}
\ee
which generalises  the classical solution \rf{rl}.
Note that $C$ is fixed to be the $\omega$-independent value between 
1/2 and 1 given by \eq{laws}. This will play a crucial role in the following.

It is interesting  to note that if we  straightforwardly insert \eq{laws} into the action \rf{aux}, it  results in a  Polyakov-like expression 
\be
S= \frac{CK_0} 2 \int \d^2 \omega \sqrt{\rho} \rho^{ab}
\partial_a X \cdot \partial_b X + K_0 \left(1-C \right) \int \d^2 \omega\sqrt{\rho}
\label{PPPa}
\ee
with  independent $\rho_{ab}$ and $X^\mu$. In the action \rf{PPPa} 
the coefficients of the quadratic in $X^\mu$ term and the volume term
obey a  certain relation. As we shall see below in Sect.~\ref{s:Pol},
this is necessary for the consistency.

Let us now discuss how $\blambda^{ab}$ depends on 
the world-sheet coordinate $\omega$.
If we choose the conformal gauge, where $\rho_{ab}$ is proportional to $\delta_{ab}$,
we have from \eq{laws} $\blambda^{ab}=C \delta ^{ab}$, \ie
$\blambda^{ab}$ is constant. This obviously satisfies \eq{Xcqsp}.
For general coordinates we expect that $\blambda^{ab}$ may depend only
on $\omega_1$ because of the cylinder geometry. We therefore obtain from
\eq{Xcqsp} the restriction
\be
\partial_1 \blambda^{11}=0,\quad \partial_1 \blambda^{12}=0 ,
\ee
so $\blambda^{11}$ and $\blambda^{12}=\blambda^{21}$ have to be
$\omega_1$-independent.
Because $\det \blambda^{ab}=C^2$, we conclude that $\blambda^{22}$
is also $\omega_1$-independent. Thus $\blambda^{ab}$ is constant.

As is shown in Sect.~\ref{s:s.p.} below, both $\brho_{ab}$ and $\blambda^{ab}$
are in fact diagonal as a consequence of the diagonal form \rf{Xcla} of the classical
solution and \eq{rhoXqsp}. However, $\brho_{11}$ and $\brho_{22}$ 
are {\em not}\/  constant and depend on $\omega_1$
 near the boundary in a nontrivial way in order that 
 the boundary conditions are satisfied (as
is discussed in Appendix~\ref{appA}).
Equation~\rf{laws} then implies that  
$\brho_{11}$ and $\brho_{22}$ will have the same $\omega_1$ dependence 
if $\blambda^{11}$ and $\blambda^{22}$ are constant since we have 
\be
\blambda^{11} \brho_{11} =\blambda^{22} \brho_{22}.
\label{gconf}
\ee

\section{Saddle-point solution at large \boldmath{$d$}\label{s:s.p.}}

To compute $\brho_{ab}$ from \eq{rhoXqsp}, we note that
\be
\left\langle  \partial_a X_{\rm q} \cdot \partial_b X _{\rm q}  \right\rangle 
=\frac d{K_0} \frac{\delta}{\delta \lambda^{ab}(\omega)}
\tr \log\left[\frac1{\sqrt{\rho}}
\left(-\partial_c\lambda^{cd}\partial_d\right)\right] .
\ee
As we have shown in the previous section, the saddle-point
value $\blambda^{ab}$ is 
$\omega$-independent. We have therefore a weaker relation
\bea
\int \d^2 \omega\, \left\langle  \partial_a X_{\rm q} \cdot \partial_b X _{\rm q}  \right\rangle 
&=&\frac d{K_0} \frac\partial{\partial \blambda^{ab}}
\tr \log\left[\frac1{\sqrt{\brho}}
\left(-\partial_c\blambda^{cd}\partial_d\right)\right] .
\label{39}
\eea

Using the proper-time regularisation~\rf{5}, we write
for the given world-sheet coordinates explicitly
\bea
\lefteqn{
\tr \log\left[\frac1{\sqrt{\brho}}
\left(-\partial_a\blambda^{ab}\partial_b\right)\right]}\non &&
=-\int_{a^2}^\infty \frac{\d\tau}\tau
\sum_{m=-\infty}^{+\infty}\sum_{n=1}^{+\infty}
 \exp\left\{- \frac{\tau}{\sqrt{\brho}}\left[
\blambda^{11} \left(\frac{\pi n}{\omega_L } \right)^2\!\!
+\blambda^{22} \left(\frac{2\pi m}{\omega_\beta } \right)^2\!\! +(\blambda^{12}+\blambda^{21}) \left(\frac{\pi n}{\omega_L } \right)\left(\frac{2\pi m}{\omega_\beta } \right)\right]\right\} .\non &&
\label{40}
\eea
We have substituted here a constant value of $\sqrt{\brho}$ because $\brho_{ab}$,
as is already pointed out (see Appendix~\ref{appA}), depends on $\omega_1$ only 
near the boundary, and the contribution from such a region will be suppressed 
in the closed string channel as
$\beta/L$. Below we will present formulas which are valid also 
for $\omega_1$-dependent $\brho_{11}$ and $\brho_{22}$, and where  this phenomenon can be explicitly observed.

The right-hand side of \eq{40} can be differentiated with respect to $\blambda^{ab}$.
Acting with  $\partial/\partial \blambda^{12}$ we find
\bea
\int \d^2\omega \left\langle  \partial_1 X_{\rm q} \cdot \partial_2 X _{\rm q}  \right\rangle 
&=&\frac{d}{K_0}\sum_{m,n}
\frac{ \left(\frac{\pi n}{\omega_L } \right)\left(\frac{2\pi m}{\omega_\beta } \right)}{\blambda^{11} \left(\frac{\pi n}{\omega_L } \right)^2\!\!+\blambda^{22} \left(\frac{2\pi m}{\omega_\beta } \right)^2\!\! +2 \blambda^{12} \left(\frac{\pi n}{\omega_L } \right)\left(\frac{2\pi m}{\omega_\beta } \right)} \non &&\times
 \exp\left\{-\frac{ a^2}{\sqrt{\brho}}\left[
\blambda^{11} \left(\frac{\pi n}{\omega_L } \right)^2\!\!+\blambda^{22} \left(\frac{2\pi m}{\omega_\beta } \right)^2\!\! +2 \blambda^{12} \left(\frac{\pi n}{\omega_L } \right)\left(\frac{2\pi m}{\omega_\beta } \right)\right]\right\},\non &&
\label{41}
\eea
where we have substituted $\blambda^{21}=\blambda^{12}$.
From \eq{41} it follows that 
\be 
\brho_{12}=0, \quad\blambda^{12}=0
\label{42}
\ee
is a solution. That it is the correct solution  can be shown order by order of the semiclassical expansion
in $d/K_0$, starting from the classical solution \rf{lacla} and using \eq{laws}. 
The reason for \rf{42} can be traced to  the diagonal form~\rf{Xcla} of the classical solution. We thus conclude that $\brho_{ab}$ and $\blambda^{ab}$ are diagonal.

For  diagonal and in general $\omega$-dependent
$\brho_{ab}$ and constant $\blambda^{ab}$ we have
\be
\tr  \log \left[\frac1{\sqrt{\brho_{11}\brho_{22}}}\left (-
 \blambda^{11}\partial_1 ^2
- \blambda^{22}\partial _2^2\right)\right] =  
-\frac{\int\d^2 \omega \sqrt{\brho_{11}\brho_{22}}}{\sqrt{\blambda^{11}\blambda^{22}}}
\Lambda^2 +
\frac{\beta\Lambda}{\sqrt{\blambda^{22}}}
+2\log
\eta \left(\frac \i2\sqrt{ \frac{ \blambda^{11}}{\blambda^{22}}}
\frac{\omega_\beta}{\omega_L}\right) ,
\label{trlog}
\ee
where the quadratic and linear divergences are as they should be 
for the proper-time regularisation. 
The finite term is given as usual~\cite{DF83} by the Dedekind eta-function.
Equation \rf{trlog} coincides with the trace log of the 2d Laplacian for the cylinder,
extracted from the general formula \cite{Alv83}.

To avoid confusion, we point out that the boundary divergence in \eq{trlog} 
(given by the second term on the right-hand side) is linked to the bulk
divergence (given by the first term on the right-hand side).
No contradiction with the open-closed string duality emerges in 
this case in contrast to Ref.~\cite{LW04}, where it was argued that the boundary
term is ruled out by open-closed string duality. In the  so-called  analytic regularisation
employed in \cite{LW04} one effectively is using $\Lambda=0$, and in that 
case the boundary term indeed vanishes.

We shall concentrate on the closed-string sector, where $L\gg\beta$
(\ie a long cylinder), and in this case the second term
on the right-hand side of \eq{trlog} can be neglected.
We then use the modular transformation of the $\eta$-function
\be
\eta\left(\frac{\i\tau }{2 }\right)=\sqrt{\frac{2}{\tau}}
\eta\left(\frac{2\i} \tau \right)
\label{modular}
\ee
and the asymptote 
\be 
\eta\left(\frac {\i \tau}{2}\right) \to \e^{-\pi \tau/24}
\label{etaexpa}
\ee
to get%
\footnote{In \eq{dec} the sign of the first term on the right-hand side
is negative when  the proper-time regularisation, but it may be positive for 
other  regularisations. For instance,  cutting of the mode expansion at some
maximal mode number leads to a positive term as shown in Ref.~\cite{AM15a}.}
\bea
\tr  \log \left[\frac1{\sqrt{\brho_{11}\brho_{22}}}\left (-
 \blambda^{11}\partial_1 ^2
- \blambda^{22}\partial _2^2\right)\right] =  
 -\frac{\int\d^2 \omega \sqrt{\brho_{11}\brho_{22}}}{\sqrt{\blambda^{11}\blambda^{22}}}
 \Lambda^2 -\frac{\pi}{3}\sqrt{\frac{\blambda^{22}}{\blambda^{11}}}
\frac{\omega_L}{\omega_\beta}.
\label{dec}
\eea

Substituting the regularised trace log from \eq{dec} into \eq{39},
 we finally  obtain
\bea
\frac{1}{\omega_\beta\omega_L} \int \d^2\omega
\brho_{11} &=&
\frac{L^2}{\omega_L^2}+\frac{\pi d}{6K_0}
\sqrt{\frac{\blambda^{22}}{(\blambda^{11})^3}}
\frac1{\omega_\beta^2}+\frac{d\Lambda^2}{2K_0}
\frac{ \int \d^2\omega\sqrt{\brho_{11}\brho_{22}}}{\sqrt{(\blambda^{11})^3\blambda^{22}}},\\ 
\frac{1}{\omega_\beta\omega_L} \int \d^2\omega
\brho_{22} &=&
\frac{\beta^2}{\omega_\beta^2}-\frac{\pi d}{6K_0\sqrt{\blambda^{11}\blambda^{22}}}
\frac1{\omega_\beta^2}+\frac{d\Lambda^2}{2K_0}
\frac{ \int \d^2\omega\sqrt{\brho_{11}\brho_{22}}}{\sqrt{\blambda^{11}(\blambda^{22})^3}} .
\eea

To solve these equations, we substitute
\be
\blambda^{11}=C \sqrt{\frac{\brho_{22}}{\brho_{11}}}, \quad
\blambda^{22}=C \sqrt{\frac{\brho_{11}}{\brho_{22}}}
\ee
as it follows
from \eq{laws} for diagonal $\brho_{ab}$ and use the already mentioned 
fact that $\brho_{11}$ and $\brho_{22}$ have the same $\omega$-dependence
(see \eq{gconf}). We then find the following solution
 \bea
\frac{1}{\omega_\beta\omega_L} \int \d^2\omega
\brho_{11}&=&\frac{L^2}{\omega_L^2}
\frac{\left(\beta^2-\frac{\beta_0^2}{2C}\right)}{\left(\beta^2-\frac{\beta_0^2}{C}\right)}
\frac{C}{2C-1},\non
\frac{1}{\omega_\beta\omega_L} \int \d^2\omega
\brho_{22}&=&\frac{1}{\omega_\beta^2}\left(\beta^2-\frac{\beta_0^2}{2C}\right)
\frac{C}{2C-1}
\label{rhogen}
\eea
and
\bea
\blambda^{11}&=& C\frac{\omega_L}{\omega_\beta L}\sqrt{\beta^2-\beta_0^2/C} ,\non
\blambda^{22}&=& C\frac{\omega_\beta L}{\omega_L}\frac1{\sqrt{\beta^2-\beta_0^2/C}} 
\label{lambdagen}
\eea
with
\be
\beta_0^2=\frac{\pi d}{3K_0}.
\ee

It should be noted that the same solution can be obtained by a straightforward minimisation of the effective action~\rf{aux1} with  \eq{dec} inserted for the trace log
and assuming that $\brho_{ab}$ and $\blambda^{ab}$ are diagonal and
constant:
\bea
S_{\rm eff}&=&\frac{K_0}2 \left( \blambda^{11}\frac{L^2\omega_\beta}{\omega_L} +
\blambda^{22}\frac{\beta^2 \omega_L}{\omega_\beta} +2\int\d^2\omega\sqrt{\brho_{11}\brho_{22}}-
\blambda^{11} \int\d^2\omega\,\brho_{11} -\blambda^{22}\int\d^2\omega\,\brho_{22}
\right) \non &&
 -\frac{\pi d}{6}\sqrt{\frac{\blambda^{22}}{\blambda^{11}}}
\frac{\omega_L}{\omega_\beta}-\frac{d\int\d^2\omega\sqrt{\brho_{11}\brho_{22}}}{2\sqrt{\blambda^{11}\blambda^{22}}}
 \Lambda^2.
\label{Sec}
\eea
This simply repeats the original Alvarez computation except that we start from  
an arbitrary  $\omega_\beta\times \omega_L$ 
rectangle in the parameter space and use the proper-time regularisation rather 
than the zeta-function regularisation.
We reproduce the results \cite{Alv81}, when $\omega_L=L$, $\omega_\beta=\beta$
and $\Lambda=0$ as it is  when using the zeta-function regularisation.
However, we emphasize once again that the more cumbersome approach we have 
used by solving Eqs.~\rf{spp} leads to  the solution \rf{rhogen} -- \rf{lambdagen} 
without  invoking the assumption that $\brho_{ab}$ 
and $\blambda^{ab}$ are diagonal and constant.


\section{The lattice-like scaling limit\label{s:ge}}

Substituting the solution~\rf{rhogen} -- \rf{lambdagen} into 
\eq{Sec}, we obtain
\be
S_{\rm eff}^{\rm s.p.} = K_0 C L \sqrt{\beta^2-\beta_0^2/C}
\label{Sfin}
\ee
for the saddle-point value of the effective action.
Further, we find that the average area of a surface which appears 
in the path integral is 
\be
{\cal A}=\langle Area \rangle = \int \d^2\omega\LA \sqrt{\det\rho_{ab}}\RA=
\int \d^2 \omega\,\sqrt{\brho_{11} \brho_{22}} =
L\frac{\left(\beta^2-{\beta_0^2}/{2C}\right)}{\sqrt{\beta^2-{\beta_0^2}/C}}\frac{C}{(2C-1)}.
\label{72}
\ee

Formulas \rf{Sfin} and \rf{72} are our main results, valid for $L \gg \beta$ in the 
mean-field approximation. Let us now discuss the physical implications of 
these formulas. 

Firstly, formula \rf{laws} for the constant $C$ (which plays the same 
role as $\alpha$ in our discussion of the random walk) shows that the bare
string tension $K_0$ needs to be renormalised in order for $C$ to remain real.
Also, $C$ is clearly constraint to take values between 1/2 and 1.
Secondly, all calculations are done with a proper time cutoff $a \sim 1/\Lambda$,
which as in the random walk case can be though of as shortest distance one
can measure in target space. Thus it is questionable if it makes sense to consider
a $\beta < a$, i.e. it does probably not make sense to enter the regime
where $S_{\rm eff}$ ceases to be real. 

At first glance it seems impossible to obtain a finite $S_{\rm eff}$ by renormalising
$K_0$ in \rf{Sfin}, since $K_0$ is of order $\Lambda^2$. However, let us try to 
imitate as closely as possible the calculation of the two-point function of the 
string by choosing, for a fixed cutoff $a$ or $\Lambda$, $\beta$ as small as 
possible without  entering into the tachyonic regime of   $S_{\rm eff}$. Thus we 
choose
\beq\label{janx10}
\beta_{\rm min}^2 = 2 \beta_0^2 \frac{K_0}{ 2d \Lambda^2}=
\frac{\pi}{3} \; \frac{1}{\Lambda^2} =  \frac{1}{3} \, (2\pi a)^2.
\eeq
This choice ensures that $\beta_{\rm min}^2 > \beta_0^2/C$ for all values 
of $K_0 > 2d \Lambda^2$ and that $\beta_0^2/C \to \beta_{\rm min}^2$ for 
$K_0 \to 2d \Lambda^2$. With this choice we have 
\beq\label{janx11}
S_{\rm eff} =  
\sqrt{\frac{\pi}{3}} \; \frac{K_0 CL}{\Lambda} \; \sqrt{2C-1}.
\eeq
Only if $ \sqrt{2C-1} \sim 1/\Lambda$ can we obtain a finite limit for 
$\Lambda \to \infty$.
Thus we are forced to renormalise $K_0$ as follows
\beq\label{janx12}
K_0 = 2d \Lambda^2 + f \frac{M_{\rm ph}^4}{\Lambda^2},
\quad f= \frac{18 d}{\pi^2}. 
\eeq

With this renormalisation we find 
\beq\label{janx12a}
S_{\rm eff} = d M_{\rm ph} L.
\eeq
Since the partition function in this case has the interpretation as a kind
of two-point function for a string propagating a distance $L$, we have 
the following leading $L$ behavior of the two-point function
\beq\label{janx13}
Z(L) \sim \e^{-S_{\rm eff}}  = \e^{-d M_{\rm ph} L+{\cal O}(\log L)} ,
\eeq
where the mass is a tunable parameter.
We note that the situation is very similar to the situation for the free particle.
In that case we had the classical value $\alpha =1$ and a semiclassical 
expansion in $1/m_0$ which interpolated between $\alpha=1$ and the 
quantum value $\alpha = 2/3$. Here we have the classical value $C=1$ and
a semiclassical expansion in $1/K_0$, which interpolates between $C=1$ and
the quantum value $C=1/2$.

In the scaling limit \rf{janx12} we can calculate the average area 
$ \langle Area \rangle = {\cal A}$ of a 
surface using \rf{72}:
\beq\label{janx14}
 {\cal A}\propto \frac{L}{M_{\rm ph}^3 a^2}.
\eeq
It diverges. If we view the surface as made from $n_{\cal A}$ building blocks
of size $a^2$, we find 
\beq\label{janx15}
n_{\cal A} \propto \frac{1}{(M_{\rm ph} L)^3} \; n_L^4,\quad n_L =  \frac{L}{a},
\eeq
telling us that the Hausdorff dimension of the surface is $d_H=4$ since 
$n_L = L/a$ is a typical linear extent of the surface measured in units of the 
cutoff $a$.

Let us finally turn to the situation where $L \gg \beta \gg a$. In this case we 
have a real extended minimal surface of area $A_{\rm min}=L \times \beta$,
around which the string fluctuates. In this case we find from \rf{72} that 
\beq\label{janx16}
{\cal A} \propto  \frac{A_{\rm min}}{M_{\rm ph}^2 a^2}.
\eeq
Again this can be written in terms of building blocks as
\beq\label{janx17}
n_{\cal A} \propto \frac{1}{M_{\rm ph}^2 A_{\rm min}} \; n_{A_{\rm min}}^2 ,\quad 
n_{A_{\rm min}} =  \frac{A_{\rm min}}{a^2},
\eeq
 showing that the Hausdorff dimension of the surface is still four for this
 kind of surfaces. 
 
 Let us now discuss what we define as the physical string tension.
 With the given boundary conditions the string extends over the minimal
 area $A_{\rm min}$ and we write the partition function as 
 \beq\label{janx20}
 Z(K_0, L, \beta) = \e^{-S_{\rm eff} (K_0,L,\beta)} = \e^{-K_{\rm ph} A_{\rm min} +
 {\cal O}(L,\beta)}.
 \eeq
 This is precisely the way one would define the physical (renormalised) string tension 
 in a gauge theory, with $L,\beta$ being the side lengths of a Wilson loop and 
 $L,\beta \gg a$, where $a$ is the lattice link length. This is also the way the 
 physical string tension is defined in lattice string theories like HLS and DT.
 Let us rewrite \rf{janx12} as 
 \beq\label{janx21}
 K_0= 2d \Lambda^2 + \frac{\tilde{K}_{\rm ph}^2}{2d \Lambda^2},
 \eeq
 very similar to the relation between the bare mass $m_0$ and the 
 renormalised mass $m_{\rm ph}$. From the explicit form of  $S_{\rm eff}$
 given in \rf{72} we have 
 \beq\label{janx22}
 K_{\rm ph} = K_0 C = d \Lambda^2 + \frac{1}{2}\tilde{K}_{\rm ph}+ 
{\cal O}(1/\Lambda^2).
 \eeq
 Thus the physical string tension as defined above diverges as the cutoff
 $\Lambda$ is taken to infinity. However, the first correction is finite and 
 behaves as we would have liked $K_{\rm ph}$ to behave, namely as 
 $\tilde{K}_{\rm ph} \propto d M_{\rm ph}^2$.  
   
We have encountered a situation identical to the one met in HLS and DT: it is
possible by renormalising the coupling constant to define a two-point 
function with a positive, finite mass. The Hausdorff dimension of 
the ensemble of surfaces is $d_H =4$,  but then the effective string tension 
defined as in \rf{janx20} will be infinite. In addition the relation \rf{janx22}
is {\it precisely}\/ the relation one finds in the lattice string theories. To make 
things clear, let us rephrase our scaling relations in dimensionless units like it 
is done in the lattice theories. Denote
\beq\label{janx23} 
K_0 a^2=\mu,\quad  d/2\pi =\mu_c,\quad K_{\rm ph}a^2 =
{\cal K}, \quad M_{ph} a^2 = {\cal M}.
\eeq
Then the renormalisation we have encountered (Eqs.\ \rf{janx12} and \rf{janx22})
can be rewritten as 
\beq\label{janx24}
{\cal M}(\mu) = c_1 (\mu-\mu_c)^{1/4}, \quad {\cal K}(\mu) = {\cal K}(\mu_c)+
c_2 (\mu-\mu_c)^{1/2}, \quad {\cal K}(\mu_c)= \mu_c/2 >0.
\eeq
These are the scaling relations obtained in lattice string theory and we have now 
reproduced them by a standard continuum mean-field calculation.

\section{Scaling to the standard string theory limit\label{standardS}}

The scaling limit of the previous section was essentially particle-like,
because the string tension has remained infinite.
Remarkably, it is possible to have yet another scaling behavior which is 
string-like and where the string tension is finite. 

We have made a decomposition $X^\mu = X^\mu_{\rm cl} + X^\mu_{\rm q}$, 
where the parameters $L$ and $\beta$ refer to the ``background''  field 
$X^\mu_{\rm cl}$. In standard quantum field theory we usually have to 
perform a renormalisation of the background field to obtain a finite effective 
action. It is possible to do the same here by scaling 
\beq\label{jany5}
X_{\rm cl}^\mu = Z^{1/2} X_R^\mu, \quad Z =(2C-1)/C.
\eeq
Notice that the field renormalisation $Z$ has a standard perturbative  expansion
\beq\label{jany6}
Z = 1 - \frac{d\Lambda^2}{2K_0} + {\cal O}(  K_0^{-2})
\eeq
in terms of the coupling constant $K_0^{-1} $, which in perturbation  theory 
is always assumed to be small, even compared to the cutoff. 

However, in the limit 
$C \to 1/2$ it has dramatic effects since,
working with renormalised lengths $L_R$ and $\beta_R$ defined as in \rf{jany5}:
\be
L_R=\sqrt{ \frac{C}{2C-1}}\; L,\qquad 
\beta_R=\sqrt{ \frac{C}{2C-1}} \;\beta,
\label{LRbetaR}
\ee
we now obtain for the effective action
\beq\label{jany7}
S_{\rm eff} = K_R \; L_R \sqrt{ \beta_R^2 - \frac{\pi d}{3 K_R}},\quad K_R = 
K_0 (2C-1)\equiv \tilde{K}_{\rm ph}.
\eeq
The renormalised coupling constant $K_R$ indeed makes $S_{\rm eff}$ 
finite and is identical to the $\tilde{K}_{\rm ph}$ defined in \rf{janx21}.
 If we view $L_R$ and $\beta_R$ as representing physical distances, \rf{jany7}
 tells us that we indeed have a renormalised, finite  string tension $\tilde{K}_{\rm ph}$ 
 in the scaling limit. In fact \rf{jany7} is identical to the continuum string
theory formula.

The ``price'' we pay for this rescaling of lengths is that 
we have introduced a tachyon in the theory. Before rescaling we argued that
the negative term under the square root was of the order of the cutoff $a^2$ and 
there was thus no compelling reason to view it as responsible for a tachyon.
However, now it has become finite and in fact it is precisely (minus) the closed bosonic 
string tachyon mass squared:
\beq\label{jany8}
M_{\rm tachyon}^2 = \frac{ \pi d}{3  \tilde{K}_{\rm ph}}.
\eeq
Looking at \rf{jany7} there is no compelling reason why $\beta_R$ could not be smaller than $M_{\rm tachyon}$. However, let us write \rf{jany5} in the following form
\beq\label{jany8a}
\frac{\beta}{2\pi a} = \sqrt{\frac{1}{\pi d}}  \; \sqrt{K_R} \beta_R.
\eeq
Thus, insisting that  $\beta/2\pi a > 1$, since $a$ plays the role of a cutoff 
distance in target space, implies that $\beta_R^2 > \pi d /K_R$, i.e.\ we are
outside the  tachyon region of  \rf{jany7}. Being deep into to tachyonic region,
i.e.\ having $\beta_R \ll M_{\rm tachyon}/K_R$ means that originally $\beta \ll a$,
clearly a situation which is strange starting out for instance in a hypercubic lattice
theory with lattice spacing $a$.

The background field renormalisation we have performed in the string 
case is very similar to the one we made for the particle, and  we can 
give it the same interpretation: the background field renormalisation is such that 
the average area ${\cal A}$ is finite, as one would define it to be if we 
considered a theory of two-dimensional gravity coupled to some matter fields.
In fact, if we insert the scaling \rf{jany5} for $X^\mu$ and \rf{jany7} for $K_0$ in
the expression \rf{72} for ${\cal A}$, we obtain
\beq\label{jany9}
{\cal A} = L_R \;
\frac{\left(\beta_R^2-\frac{\pi d}{6 K_R}\right)}{\sqrt{\beta_R^2-\frac{\pi d}{3 K_R}}},
\eeq
which  is cutoff independent and thus finite when the cutoff is removed.
The area is simply the minimal area for $\beta_R\gg M_{\rm tachyon}/K_R$ and
diverges when $\beta_R \to M_{\rm tachyon}/K_R$.

One may wonder  if it is possible to have a continuum theory 
when $L,\beta \sim a$, \ie of the order of the cutoff. Of course  it is not  in general. 
But  in the scaling limit  \rf{janx21}, where 
$(2C-1)\to \widetilde K_{\rm ph}/2d\Lambda^2$, 
the actual cutoff  is $\sim a/\sqrt[4]{\brho}\propto a\sqrt{2C-1}$, which is
much smaller than $a$. After the renormalisation \rf{LRbetaR} the cutoff becomes
$\sim a$ in the units,  where the ``physical'' distances $L_R$ and $\beta_R$
are finite, that is still much smaller than the distances. 

This phenomenon can be explicitly seen within the mode expansion,
quite similarly to what is discussed in Sect.~\ref{RW} for the relativistic
particle.
The exponent of the cutting factor [like in \eq{41}] at the saddle point  is
\begin{eqnarray}
\sum_{m,n} \frac{a^2}{\sqrt{\bar\rho}}
\left[ \bar\lambda^{11}\left(\frac{\pi n}{\omega_L}\right)^2 +\bar\lambda^{22}
\left(\frac{2\pi m}{\omega_\beta}\right)^2 \right]  
&=&  
\sum_{m,n} a^2 (2C-1) \left[ \left(\frac{\pi n}{L}\right)^2 +
\Big(\frac{2\pi m}{\sqrt{\beta^2-\beta_0^2/C}}\Big)^2 \right] \nonumber \\ 
&=&
\sum_{m,n} a^2 C \left[ \left(\frac{\pi n}{L_R}\right)^2 +
\left(\frac{2\pi m}{\sqrt{\beta_R^2-\pi d/3K_R}}\right)^2 \right].
\end{eqnarray}
So the modes are cut off  at 
$n_{max}\sim  a^{-1}{L_R}$, $m_{max}\sim
a^{-1}\sqrt{\beta_R^2-\pi d/3K_R} $.
These numbers are as large as usual ($\sim a^{-1}$) also in the 
scaling limit described in this section.

\section{Polyakov versus Nambu-Goto formulations\label{s:Pol}}

It is natural to ask if it  is  possible to reproduce the above results using the Polyakov formulation of string theory. 

Let us rewrite the Nambu-Goto action as  
\be\label{jany1}
S= (1-\alpha) K_0 \int \d^2\omega\,\sqrt{ g} 
+\frac{\alpha K_0}2 \int \d^2\omega\,  \sqrt{g} g^{ab}  \partial_a X \cdot \partial_b X, 
\ee
where $\alpha$ is a constant and where  $g_{ab}$ is the induced metric 
\be\label{jany2}
g_{ab}=\partial_a X \cdot \partial_b X.
\ee

Let us now consider $X^\mu$ and $g_{ab}$ as independent in \rf{jany1}.
We then have the Polyakov formulation of string theory. 
Integrating over quantum fluctuations of $X^\mu$, we arrive at the 
following effective action for $g_{ab}$:
\be
S_{\rm eff}= (1-\alpha) K_0 \int \d^2\omega\,\sqrt{ g} 
+\frac{\alpha K_0}2 \int \d^2\omega\,  \sqrt{g} g^{ab}  \partial_a X_{\rm cl } \cdot \partial_b X_{\rm cl } 
+ \frac d2 \tr \log \left( -\frac\alpha {\sqrt{g} }  \partial _a  \sqrt{g} g^{ab}  \partial_b  \right).
\label{auxPol}
\ee
The invariance of the measure in the path integral over $g_{ab}$ under a shift results 
in the Schwinger-Dyson equation
\be
\LA  g^{ab}\left(1-\alpha -\frac {d}{2K_0\sqrt{g}}
\langle \omega | \e^{a^2 \alpha {\Delta}}|\omega\rangle\right)\RA=0.
\label{S-DPol}
\ee
Using the Seeley expansion (in the bulk)
\be
\frac1{\sqrt{g}}\langle \omega | \e^{a^2  {\Delta}}|\omega\rangle =
\frac{1}{2\pi a^2} +\frac1{24 \pi}  R
\ee
we find this equation is consistent if $\alpha$ satisfies
\be
K_0 \left(1-\alpha \right) -\frac {d \Lambda^2}{2 \alpha}=m^2,
\label{foral}
\ee
where $m^2$ is finite, \ie if the quadratic divergence cancels.
Solving \eq{foral} for $m^2\ll K_0 \sim \Lambda^2$, we find
\be
\alpha = C= \frac 12 +\sqrt{\frac 14 - \frac {d \Lambda^2}{2K_0}}
\label{al=C}
\ee
which is already familiar from the analysis where we used  the 
Nambu-Goto action.

For the value \rf{al=C} of $\alpha$, the $\Lambda^2$ term in the
action \rf{auxPol} vanishes, so the action looks like the one obtainable  
using  the zeta-function regularisation where $\Lambda=0$ and $C=1$. 
The equation for  the Liouville field $\varphi$ (which appears  in the 
conformal gauge $\rho_{ab}=\e^{\varphi}\delta_{ab}$) is then the standard 
Liouville equation. In the string-like scaling limit  the constant  
$m^2$ in \rf{foral} is multiplied by $(2C-1)$ and the Liouville 
equation becomes a free field equation.

Thus the action \rf{auxPol} is consistent for $\alpha=1$ only for analytic
regularisations with $\Lambda=0$. Otherwise, we have to add the
non-vanishing first term. 
The Nambu-Goto formulation remarkably leads to the consistent
action, as was shown in \eq{PPPa} above.

A few comments regarding the Polyakov formulation are in order. In the conformal gauge there appears the usual ghost determinant, which can be neglected at large $d$.
Nevertheless, reparametrisations of the boundary remain essential and for the case
of the cylinder the path integral over the reparametrisations (or, equivalently, over
the boundary value of $\varphi$) reduces to an integration over the modular
parameter $\omega_\beta/\omega_L$. The latter integral can be calculated at
large $d$ again by the saddle-point method which implies a minimisation
with respect to $\omega_\beta/\omega_L$. This is in contrast to the
Nambu-Goto formulation, where $\omega_\beta/\omega_L$ was arbitrary.

The fact that $\sqrt{g}$ enters the action \rf{auxPol} linearly allows us to compute
the ground state energy. Fixing the conformal gauge, we find at the saddle point with respect to $\varphi$ for our cylinder
\be
S_{\rm eff}=\frac{K_0 C}2 \left( L^2 \frac{\omega_\beta}{\omega_L}+
\beta^2 \frac{\omega_L}{\omega_\beta}\right) -\frac{\pi d}6 
 \frac{\omega_L}{\omega_\beta}.
\label{ac}
\ee
Notice that the bulk value of $g_{ab}$ does not enter \eq{ac}.

Minimising \rf{ac} with respect to ${\omega_L}/{\omega_\beta}$, we get
\be
E_0=  K_0 C\sqrt{\beta^2 -\frac{\pi d}{3CK_0}}
\label{KKK}
\ee
that is the same as \eq{Sfin}
for the Nambu-Goto formulation. For $C=1$ we reproduce
the results~\cite{Mak11b} obtained for the zeta-function regularisation.

The mean area can be found by differentiating the partition function with respect to
$K_0$:
\be
 - K_0 \frac \partial {\partial K_0} \log Z = K_{\rm ph} \LA Area \RA,
\ee
where the string tension $K_{\rm ph}=C K_0$ from \eq{KKK},
as in \eq{janx22}. Differentiating we find
\be
 \LA Area \RA =
L \frac{(\beta^2-\beta_0 ^2 /2C)}{\sqrt{\beta^2-\beta_0 ^2 /C}}
\frac{C}{(2C-1)},
\ee
which is the same as \eq{72}  for the Nambu-Goto formulation.

However, it is not so clear how to link $g_{ab}$ to the induced metric.
They are only related by the boundary condition, stating
they are the same at the boundary~\cite{Alv83,DOP84}.

\section{Exited states\label{s:exa}}

Masses of exited states can be extracted from the next terms in the expansion
of the $\eta$-function. Using \eq{modular} with 
$\tau =\sqrt{\blambda^{11}/\blambda^{22}}\omega_\beta/\omega_L
=L^{-1}\sqrt{\beta^2-\beta_0^2/C}$,
we expand it in the closed-string sector as
\be 
\eta\left(\frac {2\i}{ \tau}\right) ^{-d}= \e^{d\pi/6 \tau} 
\prod_{n=1}^\infty \left(1-\e^{-4\pi n/ \tau} \right)^{-d}
=\e^{d\pi/6 \tau}
\sum_{N=0}^\infty d_N \e^{-4\pi  N/ \tau},
\label{etaexp1}
\ee
where $d_N$ are the level occupation numbers. 
Repeating the above computation, we obtain for the spectrum at level $N$:
\be
E_N=K_{\rm R} \sqrt{\beta^2_{\rm R} +\frac{2\pi}{K_{\rm R}}\left(4N-\frac{d}6\right)}.
\label{E_N}
\ee
For $N\sim d$ this results in a linear Regge trajectory with the ``renormalised''
Regge slop $1/8\pi K_{\rm R}=\alpha'/4$. As usual
it is four times smaller for the closed string than for an open string.

Equation~\rf{E_N} is again the usual formula for the spectrum of excited 
string states, as it follows from the zeta-function regularisation.

\section{Discussion\label{Conclusion}}

Using a mean-field continuum calculation, which we expect to be reliable in the 
large-$d$ limit, we have obtained the same result for the bosonic string 
as was originally obtained in lattice string theories (HLS or DT). In these theories
it was impossible to define a finite physical string tension in the limit where
the lattice cutoff $a$ was taken to zero. Our mean-field calculation allows us 
to trace in detail how this non-scaling arises, and it also allows us to understand 
how one in the continuum theory can perform an alternative scaling which 
reproduces some of the standard continuum results of bosonic string theory,
like formula \rf{jany7}. Rather surprisingly this scaling implies that the distances 
one considers in target space are  comparable or even much smaller than  
the cutoff $a$ one starts
out imposing. 
If one had started out with a lattice string theory like HLS where 
the path integral is performed over surfaces embedded on a hypercubic lattice
with link-lengths $a$, it clearly makes no sense to consider target space distances
less than $a$. While one in these theories {\it can}\/ define a scaling
limit for a two-point function, this scaling limit always considers distances 
much larger that $a$: when $a \to 0$
the correlation length stays finite in target space, i.e.\ it involves infinite 
many lattice spacings. 

Working in a continuum formalism, nothing prevents us from 
making an additional rescaling like \rf{jany5} of the target space, but from the 
point of the regularised theory we will, as shown explicitly by e.g. formula 
\rf{jany8a}, always be at cutoff scale $a$ for fixed rescaled distances $\beta_R$,
$L_R$ and fixed string tension $K_R$. In terms of the original variables $L,\beta$
the continuum string limit describes a Lilliputian world, which is a world where
the average area ${\cal A}$ remains finite  (as shown in formula \rf{jany9}) when 
the cutoff is removed. Having a finite ${\cal A}$ is natural from a two-dimensional 
world-sheet point of view, so our Lilliputians are naturally two-dimensional beings,
while standard lattice scaling is an enterprise only for Gulliver.
In the case of the particle this shift between the worlds of Gulliver and the 
Lilliputians is more or less an academic exercise in the sense that 
the resulting propagator was the same up to a cutoff dependent factor not 
depending on $x^\mu$. However, in the string case the Lilliputian world is the one
of  standard continuum string theory, while the Gulliver world
is one where strings are  degenerated into so-called branched polymers due to 
the non-scaling of the string tension, as described long ago in the framework of 
the HST or DT regularisation. 

From a standard field theory perspective it 
seems a little contrived  to insist that $\ell$ is finite, as one would naturally do in a
one-dimensional quantum gravity theory, since the average length of a 
world-line goes to infinite in the path integral when removing the cutoff. 
Nevertheless, as mentioned, this change of perspective has no consequences
in the case of the particle, contrary to the case of strings.
In ordinary continuum 
string calculations such a rescaling of ``distances'' $X^\mu$ which makes
the average area ${\cal A}$ finite, is usually not mentioned explicitly. 
However, it is there. Using  the conformal invariance 
of the world-sheet field theory involves a 
renormalisation of the vertex operators $e^{i p_\mu \widehat{X}^\mu(\omega)}$, i.e.\ effectively an adjustment of scales dictated by the fields $X^\mu(\omega)$.

It should also be mentioned that a finite  ${\cal A}$ is more or less the starting 
point in non-critical string theory, which can be viewed as two-dimensional 
quantum gravity coupled to matter fields with central charge $c < 1$.  In these 
theories the finite ${\cal A}$ is obtained by a renormalisation of the two-dimensional
gravitational cosmological term, i.e.\ the first term on the
right-hand side of \eq{jany1}. Some 
of our calculations can formally be extended to the region $c <1$, since this 
region, again formally, corresponds to $d < 0$. As is seen from our formulas
everything is different if $d < 0$ and one obtains completely different scaling.
Such a different scaling could well be consistent with the scaling obtained by
the DT lattice theory which for $c < 1$ provides a regularisation of two-dimensional
quantum gravity coupled to matter, and where the scaling, contrary to 
the situation for $d>0$, agrees with continuum non-critical 
string calculations. Our mean-field results might be reliable in the $d \to -\infty$ 
limit, but we have not investigated it in detail.

Our results are based on the mean-field approximation and
reproduce in the string-like scaling limit  the spectrum obtained by the canonical
quantisation. It would be interesting to pursue our approach beyond the 
mean-field approximation, accounting for fluctuations of $\rho_{ab}$ and
$\lambda^{ab}$ to next orders in  $1/d$, 
to check whether or not the spectrum changes.

\subsection*{Acknowledgments}

We are indebted to Poul Olesen, Peter Orland and Arkady Tseytlin for valuable discussions.

The authors acknowledge  support by  the ERC-Advance
grant 291092, ``Exploring the Quantum Universe'' (EQU).
Y.~M.\ thanks the Theoretical Particle Physics and Cosmology group 
at the Niels Bohr Institute for the hospitality. 

\appendix

\section{More on relativistic paths\label{AppB}}

Let us explicitly check that the solution \rf{B13} to \eq{B4} is the one which
sums up the semiclassical expansion in $1/m_0$. The proper exact solution
to the cubic \eq{B4} at large $L$ is 
\bea
\alpha(\ka)&=&
\frac{2}{3}+\frac{\left(1+i \sqrt{3}\right) \left(2-27 \ka^2
+i 3 \sqrt{3} \ka\sqrt{4-27 \ka^2}\right)^{1/3}}{2^{1/3} 6}\non &&+
\frac{\left(1-i \sqrt{3}\right)\left(2-27 \ka^2-i 3 \sqrt{3} \ka\sqrt{4-27 \ka^2}\right)^{1/3}}{2^{1/3} 6},
\eea
where $\ka=d \Lambda /2 m_0$. It has the required series expansion
\be
\alpha=1-\ka-\frac{\ka^2}{2}-\frac{5 \ka^3}{8}-\ka^4-\frac{231 \ka^5}{128}-\frac{7 \ka^6}{2} 
+{\cal O}(\ka^8),
\ee
monotonically decreases with $\ka$ 
and indeed for $\ka=\sqrt{\frac2{27}}-\delta$
\be
\alpha=\frac23+2\sqrt{\frac23}\delta +{\cal O}(\delta^2).
\ee

To explicitly compute the induced metric in the static gauge, we use
the mode expansion
\be
x_q=\sqrt{2} \sum_{n =1}^\infty  a_{n}\sin\frac{\pi n \omega}{\omega_L}.
\ee
We then obtain
\be
\LA \dot x_q ^2 \RA =\frac{2}{\omega_L}
 \sum_{n=1}^\infty \left( \frac{\pi n}{\omega_L} \right)^2  \LA a_{n}^2 \RA
\cos^2 \frac{\pi n}{\omega_L}\omega \e^{-a^2 \lambda \left( \frac{\pi n}{\omega_L}\right)^2h^{-1/2}}
 =
\frac{2}{\omega_L}\frac{d}{m_0 \lambda}
\sum_{n=1}^\infty 
 \cos^2 \frac{\pi n}{\omega_L}\omega  \e^{-a^2 \lambda \left( \frac{\pi n}{\omega_L}\right)^2h^{-1/2}}.
\label{A88p}
\ee
Replacing the sum by an integral, we find
\be
\LA \dot x_q ^2 \RA =\frac{2d}{m_0\lambda} \int_0^\infty
\d x \cos^2 x  \e^{-a^2 \lambda x^2h^{-1/2}} =
\frac{d \Lambda \sqrt[4]{h}}{m_0\lambda^{3/2}}.
\label{B14a}
\ee
This simply reproduces \eq{hhhhh} without the last term, which comes 
from the difference between the sum in \eq{A88p} and the integral in \eq{B14a}.

We have seen that nothing unexpected happens with the induced metric 
in the case of paths. In particular, we can make it constant by choosing the 
proper-time gauge \rf{gaugep}. This is in contrast to the case of surfaces, 
where the dependence of the induced metric on $\omega$ is present
to fulfill the boundary condition for the component of the metric
tensor along the boundary, as is demonstrated in Appendix~\ref{appA}.

To see how the typical paths that dominate the path integral look like,
let us compute the averaged transverse displacement squared 
$\LA x_\perp^2 \RA$. Proceeding as in \eq{A88p}, we find
\be
\LA  x_\perp ^2 \RA =\frac{2}{\omega_L}
 \sum_{n=1}^\infty  \LA a_{n}^2 \RA
\sin^2 \frac{\pi n}{\omega_L}\omega
 =
\frac{2}{\omega_L}\frac{d}{m_0 \lambda}
\sum_{n=1}^\infty 
 \frac{\sin^2 \frac{\pi n}{\omega_L}\omega }
{ \left( \frac{\pi n}{\omega_L} \right)^2} 
\label{A88pp}
\ee
which is convergent. 
At large $L$ we can replace the sum by an integral to get
\be
\LA  x_\perp ^2 \RA =\frac{2d}{\pi m_0 \lambda} \int_0^\infty \frac{\d x}{x^2}\sin^2 x \omega=\frac{d\omega}{m_0\lambda} = \frac{dL}{m_0 \sqrt{\alpha (3\alpha-2)}}
\frac {\omega}{\omega_L}
\label{trtr}
\ee
for $\omega<\omega_L/2$. In the scaling regime \rf{B140}  
it tends to a finite value 
\be
\LA  x_\perp ^2 \RA =
\frac{dL}{m_{\rm ph}} \frac {\omega}{\omega_L}
\label{tdis1}
\ee
with the transverse displacement growing
like $\sqrt{L}$, as it should for the Brownian motion in the target space.

In the other scaling regime \rf{B141}, the right hand side of \eq{trtr} vanishes
as $m_0^{-1}\sim \Lambda^{-1}$. 
However, if we renormalise the transverse coordinate in the same way as in
 \eq{janz3}, the transverse displacement would also be finite
\be
\LA  x_R{}_\perp ^2 \RA =
\frac{dL_R}{\tilde{m}_{\rm ph}} \frac {\omega}{\omega_L}
\label{tdis2}
\ee
and coinciding with \rf{tdis1}.

We can also compute the correlator at non-coinciding ``times'' $\omega_1$
and $\omega_2$.
The 1d Dirichlet Green function can be computed through the mode
expansion quite similarly to Eqs.~(\ref{A88pp}) -- (\ref{tdis2}). We  obtain
\begin{eqnarray}
\langle x^\mu_{\rm q}(\omega_1) x^\nu_{\rm q}(\omega_2) \rangle&=&
\frac{2}{\omega_L}\frac{\delta^{\mu\nu}}{m_0 \lambda}
\sum_{n=1}^\infty 
 \frac{\sin \frac{\pi n}{\omega_L}\omega_1 \sin \frac{\pi n}{\omega_L}\omega_2}
{ \left( \frac{\pi n}{\omega_L} \right)^2} \nonumber \\*
& = &\frac{\delta^{\mu\nu}L}{m_0 \sqrt{\alpha (3\alpha-2)}} \left(
\frac {\omega_1+\omega_2}{2\omega_L} -
\frac {|\omega_1-\omega_2|}{2\omega_L}
\right),
\label{Green}
\end{eqnarray}
which is valid for $\omega_i < \omega_L/2$ (otherwise 
$\omega_i \Longrightarrow \omega_L -\omega_i$) and reproduces \eq{trtr}
for $\omega_1=\omega_2$. 
It vanishes if $\omega_i=0$ (i.e.\ at the boundary), as the Dirichlet Green function
should. The first term in the brackets makes it positive.

In Eq.~(\ref{Green})  the coefficient $\frac{L}{m_0 \sqrt{\alpha (3\alpha-2)}} $ equals
either $\frac{L}{m_{\rm ph}}$ in the scaling limit \rf{B140} or
$\frac{L_R}{\widetilde m_{\rm ph}}$ in the scaling limit \rf{B141}, if we 
renormalise $x^\mu_{\rm q}$.
So the continuum  formulas are identical in both cases.

\section{Induced metric in the world-sheet coordinates \label{appA}}

Let us compute the induced metric
$\LA \partial_a X \cdot \partial_b X \RA$ in the string case
to verify its coordinate dependence.

Using the mode expansion
\be
X_q=2
 \sum_{m,n \geq0} \left( a_{mn} \cos\frac{2\pi m\omega_2}{\omega_\beta}
+b_{mn} \sin\frac{2\pi m\omega_2}{\omega_\beta}
\right) \sin\frac{\pi n \omega_1}{\omega_L},
\ee
we have explicitly for the quantum part of the induced metric
\bea
\lefteqn{
\LA \partial_1 X_q \cdot \partial_1 X_q \RA}\non
 &=&
\frac{2}{\omega_\beta \omega_L }
\sum_{m=0}^\infty \sum_{n=1}^\infty \left( \frac{\pi n}{\omega_L} \right)^2 \left[
(2-\delta_{m0} ) \LA a_{mn}^2 \RA\cos^2  
\frac{2\pi m}{\omega_\beta}\omega_2  
 +2\LA b_{mn}^2 \RA\sin^2  
\frac{2\pi m}{\omega_\beta}\omega_2  
 \right] \cos^2 \frac{\pi n}{\omega_L}\omega_1\non &=&
\frac{2}{\omega_\beta \omega_L}\frac{d}{K_0 }
\sum_{m=-\infty}^{+\infty} \sum_{n=1}^\infty \frac {\left( \frac{\pi n}{\omega_L} \right)^2}
{ \blambda^{22}\left( \frac{2\pi m}{\omega_\beta} \right)^2+ \blambda^{11}\left( \frac{\pi n}{\omega_L} \right)^2}
 \cos^2 \frac{\pi n}{\omega_L}\omega_1
\label{A88}
\eea

The sum over $m$ is convergent and easily done, while
the sum over $n$ can be substituted by an integral
as $L\to \infty$ (the closed string channel).
For the divergent part we find
\bea
\lefteqn{
\frac{d}{\pi K_0 \sqrt{(\blambda^{11})^3 \blambda^{22}}}
\int_0^\infty \d x\, \frac{x ^2}{y^2+x^2}
\cos^2 \left(\frac{x \omega_1}{\sqrt{\lambda^{11}}}\right) \e^{-\varepsilon  x^2} } \non
&&=\frac{d}{K_0 \sqrt{(\blambda^{11})^3 \blambda^{22}}}
\frac1{8 \pi \varepsilon}
\left[{1-\frac{\varepsilon\lambda^{11}}{\omega_1^2}+\e^{-\frac{\omega_1^2}{\varepsilon\blambda^{11}}} \left(2+\frac{\varepsilon\blambda^{11}}{\omega_1^2}\right)}\right],
\label{A3}
\eea
where
\be
\varepsilon =\frac{a^2}{\sqrt{\brho_{11}\brho_{22}}}=
\frac{1}{4\pi \Lambda^2\sqrt{\brho_{11}\brho_{22}}}.
\ee

For the finite part we get
\bea
\lefteqn{
\frac{d}{K_0 \sqrt{(\blambda^{11})^3 \blambda^{22}}}
\frac{1}{\pi}\int_0^\infty \d x\, x 
\left[\coth \left(  \frac{\omega_\beta x}{2\sqrt{\blambda^{22}}}
\right)-1 \right]
\cos^2 \left(\frac{x \omega_1}{\sqrt{\lambda^{11}}}\right)}\non &&=\frac{\pi d}{6 K_0 \omega_\beta^2}
\frac{\sqrt{\blambda^{22}}}{(\blambda^{11})^{3/2}} +
\frac{d}{K_0 \sqrt{\blambda^{11} \blambda^{22}}}
\frac{1}{\pi}\left[\frac{1}{8\omega_1^2}-\frac{\pi^2 \blambda^{22}}{2\omega_\beta^2\lambda^{11}}
\frac1{ \sinh^2\left(\frac{2\pi \omega_1}{\omega_\beta}\sqrt{\frac{\blambda^{22}}{\blambda^{11}}} \right)}
\right].
\label{A4}
\eea
The first term on the right-hand side is familiar from the integrated version of
Sect.~\ref{s:s.p.}.
The second term  makes the induced metric to be 
$\omega_1$-dependent.  At $\omega_1= 0$ it is equal to the first term.

Adding \rf{A3} and \rf{A4}, we finally obtain
\bea
\LA \partial_1 X_q \cdot \partial_1 X_q \RA &=&
\frac{d}{K_0 \sqrt{(\blambda^{11})^3 \blambda^{22}}}
\frac1{8 \pi \varepsilon}
\left[{1+\e^{-\frac{\omega_1^2}{\varepsilon\blambda^{11}}} \left(2+\frac{\varepsilon\blambda^{11}}{\omega_1^2}\right)}\right]
\non&&+\frac{\pi d}{6 K_0 \omega_\beta^2}
\frac{\sqrt{\blambda^{22}}}{(\blambda^{11})^{3/2}} -
\frac{\pi d}{2K_0 \omega_\beta^2}\frac{\sqrt{\blambda^{22}}}{(\blambda^{11})^{3/2}}
\frac1{ \sinh^2\left(\frac{2\pi \omega_1}{\omega_\beta}\sqrt{\frac{\blambda^{22}}{\blambda^{11}}} \right)}.
\label{A89}
\eea

We see from  \eq{A88} that the mean value equals
\bea
\frac1{\omega_\beta \omega_L}\int \d^2 \omega\LA \partial_1 X_q \cdot \partial_1 X_q \RA 
&=&\frac{d }{8\pi \varepsilon K_0 \sqrt{(\blambda^{11})^3\blambda^{22}}}
+\frac{\pi d}{6 K_0 \omega_\beta^2}
\frac{\sqrt{\blambda^{22}}}{(\blambda^{11})^{3/2}} 
\label{A90}
\eea
which coincides with the right-hand side of \eq{A89} far away from the boundary.
The fact that the induced metric is $\omega_1$-dependent near the boundary 
does not affect the mean value, because its contribution to the mean value is 
${\cal O}(1/L)$.
This $\omega_1$-dependence of the induced
metric near the boundary is specific to the cylinder (and disk) topology.
It would be missing for a torus.

Exactly at the boundary we have from \eq{A89} the twice larger value then 
the mean value \rf{A90}
\bea
\LA \partial_1 X_q \cdot \partial_1 X_q \RA \big|_B&=&
\frac{d}{4 \pi\varepsilon K_0 \sqrt{(\blambda^{11})^3\blambda^{22}}}
+\frac{\pi d}{3 K_0 \omega_\beta^2}
\frac{\sqrt{\blambda^{22}}}{(\blambda^{11})^{3/2}}
 .
\label{bc1q}
\eea
As we shall momentarily see, this guarantees for the gauge condition 
\rf{gconf} to be satisfied in the bulk.

Analogously, we find
\bea
\lefteqn{\LA \partial_2 X_q \cdot \partial_2 X_q \RA  =\frac1{\omega_\beta\omega_L}
\frac{2d}{K_0}
\sum_{m,n} \frac {\left( \frac{2\pi m}{\omega_\beta} \right)^2}
{ \blambda^{22}\left( \frac{2\pi m}{\omega_\beta} \right)^2+ \blambda^{11}\left( \frac{\pi n}{\omega_L} \right)^2}
 \sin^2 \frac{\pi n}{\omega_L}\omega_1}\non 
&&=
\frac{ \blambda^{11}}{\blambda^{22}}
\LA \partial_1 X_q \cdot \partial_1 X_q \RA -\frac1{\omega_\beta\omega_L}\frac{2d}{K_0 \blambda^{22}}
\sum_{m,n} \left[
 \frac {  \blambda^{11}\left( \frac{\pi n}{\omega_L} \right)^2}
{ \blambda^{22}\left( \frac{2\pi m}{\omega_\beta} \right)^2+ \blambda^{11}\left( \frac{\pi n}{\omega_L} \right)^2} -\sin^2 \frac{\pi n}{\omega_L}\omega_1\right] \non
&&= \frac{ \blambda^{11}}{\blambda^{22}}
\LA \partial_1 X_q \cdot \partial_1 X_q \RA- \frac{ \blambda^{11}}{\blambda^{22}}
\LA \partial_1 X_q \cdot \partial_1 X_q \RA \big|_B +
\frac{d}{K_0 \sqrt{\blambda^{11}(\blambda^{22})^3}}
\frac{1}{4\pi \varepsilon } \left(1- \e^{-\omega_1^2/\varepsilon\blambda^{11}}\right).
\label{uuuufff}
\eea
We see that at the boundary
\be
\LA \partial_2 X_q \cdot \partial_2 X_q \RA \big|_B=0
\ee
because of the boundary condition $X_q \big|_B=0$ and
because the derivative is along the boundary.

Using \eq{bc1q} and adding the classical parts, we rewrite \eq{uuuufff} as
the following relation between components of the whole induced metric:
\be
\blambda^{11} \LA \partial_1 X \cdot \partial_1 X \RA=
\blambda^{22} \LA \partial_2 X \cdot \partial_2 X \RA +
\frac{d}{4\pi \varepsilon K_0 \sqrt{\blambda^{11}\blambda^{22}}}
 \e^{-\omega_1^2/\varepsilon\blambda^{11}}.
\ee
We see that $\LA \partial_2 X \cdot \partial_2 X \RA$ equals $\LA \partial_1 X \cdot \partial_1 X \RA$ everywhere outside the 
$\varepsilon$-vicinity of the boundary, where a more careful analysis of
the second term on the right-hand side is required.

Using the mode expansion, we can also compute the transversal size of the
string. Proceeding as above, we get
\bea
\LA X_{\rm q}^2 \RA &=&
\frac{1}{\omega_\beta \omega_L}\frac{2d}{K_0 }
\sum_{m=-\infty}^{+\infty} \sum_{n=1}^\infty \frac {1}
{ \blambda^{22}\left( \frac{2\pi m}{\omega_\beta} \right)^2+ \blambda^{11}\left( \frac{\pi n}{\omega_L} \right)^2}
 \sin^2 \frac{\pi n}{\omega_L}\omega_1 \non 
 &=&
{
\frac{d}{\pi K_0 \sqrt{\blambda^{11} \blambda^{22}}}
\int_0^\infty \frac{\d x} x 
\coth \left(  \frac{\omega_\beta x}{2\sqrt{\blambda^{22}}}
\right)
\sin^2 \left(\frac{x \omega_1}{\sqrt{\blambda^{11}}}\right)} \e^{-\varepsilon x^2}.
\label{lalast}
\eea

The integral in \eq{lalast} has a logarithmic domain for $\varepsilon \ll \omega_1^2 /\blambda^{11}$
The (logarithmically) divergent part of the integral is
\be
\int_0^\infty \frac{\d x} x 
\sin^2 \left(\frac{x \omega_1}{\sqrt{\blambda^{11}}}\right) \e^{-\varepsilon x^2}=\frac{1}4
\left( \log\frac{4\omega_1^2}{\varepsilon\blambda^{11}} + \gamma_{\rm E} \right)
+{\cal O}(\varepsilon).
\ee
The finite part is
\be
\int_0^\infty \frac{\d x} x \left[
\coth \left(  \frac{\omega_\beta x}{2\sqrt{\blambda^{22}}}
\right)-1 \right]
\sin^2 \left(\frac{x \omega_1}{\sqrt{\blambda^{11}}}\right)=\frac 12 \log \sinh
\frac{2\pi \omega_1 {\sqrt{\blambda^{22}}}}{\omega_\beta{\sqrt{\blambda^{11}}}}
-\frac 12 \log 
\frac{2\pi \omega_1 {\sqrt{\blambda^{22}}}}{\omega_\beta{\sqrt{\blambda^{11}}}} ,
\ee
Finally, we obtain for large $L$
\bea
\LA X_{\rm q}^2 \RA &= &\frac{d}{K_0 C}\left\{\frac{1}{4\pi}
\left( \log\left[\frac{\omega_\beta \omega_L \sqrt{\beta^2-\beta_0^2/C}}
{\varepsilon \pi^2 C L}
\right]+ \gamma_{\rm E} \right)
+\frac{\omega_1 }{\omega_L}\frac L{\sqrt{\beta^2-\beta_0^2/C}}
\right\} .
\label{B155}
\eea
If we perform 
the above renormalisation \rf{jany5} of the length scale 
 $X_{\rm q}^2 \to (2C-1) [X_{\rm q}^2]_{\rm R}/C$, then $K_0$ in the denominator 
becomes $K_{\rm R}$:
\be
\LA [ X_{\rm q}^2]_{\rm R} \RA = \frac{d}{K_ {\rm R}}\left\{\frac{1}{4\pi}
\left( \log\left[\frac{4 \Lambda^2 }\pi\left(\beta^2_{\rm R}-\frac{\pi d}{ 6 K_{\rm R}}\right)
 \right]+ \gamma_{\rm E} \right)
+\frac{\omega_1 }{\omega_L}\frac{ L_{\rm R}}{\sqrt{\beta^2_{\rm R}-\pi d /3 K_{\rm R})}}
\right\} .
\label{mis}
\ee

The first term on the right-hand side of \eq{mis} is familiar from the open-string case.
It has the logarithmic divergence which cannot be renormalised, so it always diverges.
It is the same for the zeta-function regularisation, where the log is
replaced by $\zeta(1)=\infty$. We shall soon return to this issue.

The appearance of the second term on the right-hand side of \eq{mis} is specific to a cylinder. It comes from the modes with $m=0$ (the zero mode) and is missing
for an open string.
It looks pretty much like the one  in \eq{tdis2} for the random paths,
if we identify $\tilde{m}_R$ with the mass of the lowest string state which 
propagates the distance $L_R$.

We can also compute the whole Dirichlet propagator at different
$\omega$ and $\omega^\prime$, replacing the sum over $n$ by an
integral at large $L$ and summing over $m$ by using the formulas
\begin{eqnarray}
\int_0^\infty \frac{{\rm d} x}{a^2+x^2} \sin(bx)\sin(cx) &
=&\frac{\pi}{4a}\left( {\rm e}^{-a|b-c|} -{\rm e}^{-a(b+c)}\right) ,\\
\sum_{m=1}^\infty  \frac 1m {\rm e}^{-b m} \cos(a m)&=&
-\frac12 \log \left( 1 - 2\,{\rm e}^{-b } \cos a +{\rm e}^{-2b } \right).
\end{eqnarray}
Integrating first over $n$ and then summing over $m$, we obtain
\begin{eqnarray}
\langle X^\mu_{\rm q}(\omega) 
X^\nu_{\rm q}(\omega^\prime) \rangle &=&
\frac{2 \delta^{\mu\nu}}{\omega_\beta \omega_L K_0}
\sum_{m=-\infty}^{+\infty} \sum_{n=1}^\infty \frac { \sin \frac{\pi n}{\omega_L}\omega_1 \sin \frac{\pi n}{\omega_L}\omega_1^\prime}
{ \bar\lambda^{22}\left( \frac{2\pi m}{\omega_\beta} \right)^2+ \bar\lambda^{11}\left( \frac{\pi n}{\omega_L} \right)^2}
 \cos \frac{2\pi m}{\omega_\beta}(\omega_2-\omega_2^\prime),\nonumber \\
 &=&
\frac{\delta^{\mu\nu}}{K_0 C} \frac1{4\pi} 
\log\frac{ \left(  \cosh\frac{2\pi (\omega_1+\omega_1^\prime)}{\omega_\beta } - 
 \cos \frac{2\pi (\omega_2-\omega_2^\prime)}{\omega_\beta} \right)}
{ \left(  \cosh\frac{2\pi (\omega_1-\omega_1^\prime)}{\omega_\beta } - 
 \cos \frac{2\pi (\omega_2-\omega_2^\prime)}{\omega_\beta} \right)},
\label{DGreen}
\end{eqnarray}
where we set
\begin{equation}
\omega_L=L,\quad \omega_\beta=\sqrt{\beta^2-\beta_0^2/C}
\end{equation}
for simplicity of the formulas.

Equation~(\ref{DGreen}) represents the 2d 
(non-regularised) Dirichlet Green function for a cylinder.
If $|\omega-\omega'|\ll \omega_\beta$, (\ref{DGreen}) behaves as
\begin{equation}
 ({\rm \ref{DGreen}})  \stackrel{|\omega-\omega'|\ll \omega_\beta}\to
 -\frac{\delta^{\mu\nu}}{K_0 C}\frac1{2\pi} \log|\omega-\omega'|,
\end{equation}
i.e.\ as the ordinary Green function.

We can also compute 
$\langle \partial_a X_{\rm q} (\omega)\cdot
\partial_b X_{\rm q} (\omega^\prime)\rangle $ for 
$\omega_\beta\gg|\omega-\omega^\prime| \gg
\sqrt{\varepsilon} =a\sqrt[4]{\bar\rho}$
by differentiating (\ref{DGreen})
with respect to $\omega_a$ and $\omega_b '$. We then find
\begin{equation}
\langle \partial_1X_{\rm q}(\omega_1,\omega_2) \cdot
\partial_1 X_{\rm q}(\omega_1^\prime,\omega_2) \rangle 
\stackrel{\omega_1^\prime\to\omega_1}\to
\frac{d}{K_0 C}\frac{\pi}{2\omega_\beta^2}
\left( -\frac1{\pi^2(\omega_1-\omega_1^\prime)^2} + \frac {1}3 - \frac1{\sinh \frac{2\pi \omega_1}{\omega_\beta} }  \right)
\end{equation}
for $\omega_2^\prime=\omega_2$.
In particular, we recover this way  the $\omega_1$-dependent term in the final
part of the induced metric displayed in Eq.~(\ref{A89}).
The constant term is also reproduced but this could be a coincidence because
it is in general regularisation-dependent. 

It is not hard to compute a correlator  analogous to \rf{Green} in the string case.
Setting in \eq{DGreen} $\omega_2^\prime=\omega_2$, 
we find
\begin{equation}
\langle X^\mu_{\rm q}(\omega_1,\omega_2) 
X^\nu_{\rm q}(\omega_1^\prime,\omega_2) \rangle  
=\frac{\delta^{\mu\nu}}{2\pi K_0 C}
\left(\log \sinh
\frac{\pi (\omega_1+\omega_1^\prime)L}{\omega_L{\sqrt{\beta^2-\beta_0^2/C}}}-\log \sinh
\frac{\pi |\omega_1-\omega_1^\prime| L}{\omega_L{\sqrt{\beta^2-\beta_0^2/C}}}
\right).
\label{sGreen}
\end{equation}
At large $L$ only the zero mode (i.e. the $m=0$ modes) remains with an
exponential accuracy and we get 
\begin{eqnarray}
\langle X^\mu_{\rm q}(\omega_1,\omega_2) 
X^\nu_{\rm q}(\omega_1^\prime,\omega_2) \rangle&=&
\frac{\delta^{\mu\nu} L}{K_0 C{\sqrt{\beta^2-\beta_0^2/C}}}
\left(
\frac{\omega_1+\omega_1^\prime}{2\omega_L}-
\frac{|\omega_1-\omega_1^\prime| }{2\omega_L}
\right),
\end{eqnarray}
which is quite analogous to Eq.~(\ref{Green}) in the particle case.
In the open-string case this zero mode was absent and the 
contribution of nonzero modes coincides with the open-string result~\cite{Alv81}.

Remarkably, the log divergence, contaminating Eqs.~({\ref{B155}), (\ref{mis})
is missing in the corrlelator~\rf{sGreen}. 
It comes back if $|\omega_1-\omega_1^\prime| \lesssim
\sqrt{\varepsilon} =a\sqrt[4]{\bar\rho}$. 
This could be most probably interpreted as an effect of spikes, 
\ie very long thin pieces of surfaces of negligible area,
of longitudinal size of the cutoff at  the world-sheet.

\end{document}